\documentclass[letterpaper]{article}

\usepackage[T1]{fontenc}

\usepackage{geometry}
\usepackage{setspace}
\usepackage{graphicx}
\usepackage{float}
\newfloat{scheme}{htbp}{los}
\floatname{scheme}{Scheme}
\floatname{chart}{Chart}
\newfloat{graph}{htbp}{loh}

\usepackage{chemformula} 

\newcommand*\bmc[1]{\bm{\mathcal{#1}}}

\usepackage{amsmath,amssymb,amsfonts}
\usepackage{epsfig}
\usepackage{esint}
\usepackage{soul}
\usepackage{bm}
\usepackage{siunitx}
\graphicspath{{images/}}
\usepackage[export]{adjustbox}
\usepackage{algorithm}
\usepackage{algorithmic}
\usepackage{placeins}

\usepackage{macro}

\DeclareSIUnit{\molar}{M}

\usepackage{authblk}
\author[1,2]{Adam R. Lamson*}
\author[1]{Mohammadhossein Firouznia}
\author[1,3]{Michael J. Shelley}
\affil[1]{Center for Computational Biology, Flatiron Institute, New York, NY 10010 USA}
\affil[2]{Cluster of Excellence Physics of Life, TU Dresden, Dresden, 01307 Germany}
\affil[3]{Courant Institute, New York University, New York, NY, 10012 USA}

\title{Condensation dynamics of sticky and anchored flexible biopolymers}
\date{*Email: alamson@flatironinstitute.org}

\begin{document}
\maketitle




\begin{abstract}
Cells regulate gene expression in part by forming DNA–protein condensates in the
nucleus. While existing theories describe the equilibrium size and stability of
such condensates, their dynamics remain less understood. Here, we use
coarse-grained 3D Brownian-dynamics simulations to study how long, end-anchored
biopolymers condense over time due to transient crosslinking. By tracking how
clusters nucleate, merge, and disappear, we identify two dominant dynamical
pathways, ripening and merging, that govern the progression from an uncompacted
chain to a single condensate. We show how microscopic kinetic parameters,
protein density, and mechanical constraints shape these pathways. Using insights
from the simulations, we construct a minimal mechanistic free-energy model that
captures the observed scaling behavior. Together, these results clarify the
dynamical determinants of DNA and chromatin reorganization on timescales
relevant to gene regulation.
\end{abstract}

\maketitle

\begin{figure}[h!]
  \centering
  \adjustbox{trim={0.0\width} {0.0\height} {0.0\width} {0.0\height},clip}
  {\includegraphics[width=1\columnwidth, angle=0]{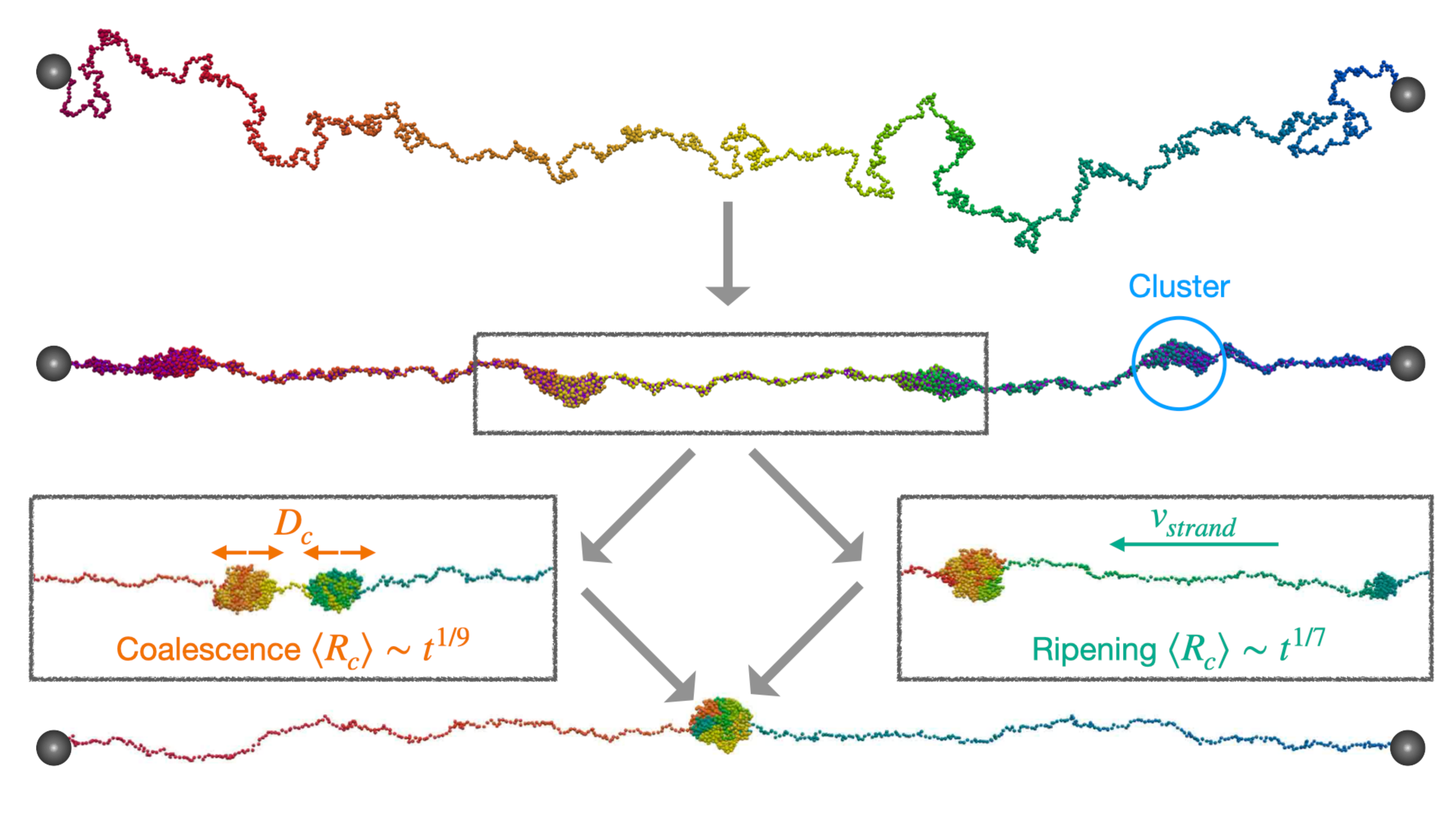} }
\end{figure}

\FloatBarrier

\section{Introduction}%
\label{sec:introduction}

Biopolymers provide the structural and functional framework for many biological
processes. From RNA and amino acid chains to cytoskeletal filaments and
chromatin, organisms rely on the precise organization of these macromolecules to
sustain cellular functions \cite{Alberts2017}. Among them, chromatin—the complex
of DNA and histone proteins within eukaryotic nuclei—is particularly remarkable.
A single human chromosome, comprising approximately $10^8$ nucleotides ($\sim
10^4 \mu$m), is intricately compacted into a nucleus of $\sim 10 \mu$m in
diameter \cite{VanHolde1989}. Recent studies have revealed that chromatin is not
merely a passive carrier of genetic information but a highly organized, dynamic
structure\cite{Krietenstein2020,Nagano2017,Amiad-Pavlov2021}. Disruptions to
chromatin organization have been linked to developmental disorders, cancer, and
other diseases \cite{Zhou2016,Gambi2025,Kloetgen2020,Fang2020,Krumm2019}. 

This organization is maintained by a diverse set of proteins that
selectively bind, interpret, and remodel chromatin, dynamically shaping genome
accessibility and activity \cite{Barral2024, Hoencamp2023,
DiGiammartino2020,Hsieh2020}. A central question in nuclear organization is how
cells establish and regulate a chromatin state that is both highly structured
and adaptable. One proposed strategy is biomolecular condensation, where
proteins transiently associate with chromatin by binding to DNA, histones, and
other nuclear factors, driving self-association and therefore compartmentalizing
or clustering of genomic regions \cite{Sabari2020, Sanulli2020}. 
This process plays a fundamental role in genome function, influencing
transcriptional regulation, DNA repair, and the establishment of epigenetic
modifications \cite{Henninger2021,Owen2023,Cho2018}.
    

The dynamic reorganization of chromatin takes place within the complex and
crowded nuclear environment, where chromatin is tethered to nuclear structures
such as the lamin fibers lining the nuclear envelope and the nucleoli
\cite{Zheng2018,Amiad-Pavlov2021,Pontvianne2016}. This raises the fundamental
question of how passive nuclear proteins, such as HP1, orchestrate the
large-scale reorganization of chromatin within confined and geometrically
constrained nuclear environments while ensuring the speed and selectivity
required for proper cellular function.

\textit{In vivo} studies of nuclear hubs, speckles, and other condensates have
demonstrated chromatin organization's importance in cellular function.
Single-molecule \textit{in vitro} experiments complement such experiments by
elucidating the polymeric and physical principles governing DNA-protein
condensation
\cite{Golfier2020,Keenen2021,Quail2021,Morin2022,Renger2022,Soranno2021}. By
manipulating DNA and chromatin segments between optically trapped beads,
experiments measure forces, molecular interactions, and protein distributions
along the polymers \cite{Wang1997,Kumar2010,Quail2021,Nguyen2022}. Furthermore,
assays have also recreated the dynamic organization of DNA using Xenopus egg
extract containing all the proteins present in nuclear
environments \cite{Sun2023}. 

Such research has revealed that protein-protein interactions within nuclear
condensates are often mediated by the intrinsically disordered regions (IDRs) of
DNA-binding proteins \cite{Michieletto2022}. These interactions, typically on
the order of $1\,k_B\,T-3\,k_B\,T$ \cite{Morin2022,Schnatwinkel2020}, are
individually weak but collectively generate sufficient force to drive chromatin
condensation \cite{Quail2021,Renger2022,Soranno2021,Zhao2024}. Additionally,
specific proteins exhibit prewetting behavior, coating DNA in a monolayer and,
at higher protein concentrations, form droplets or condensed homogenous
polymer-protein mixtures known as co-condensates \cite{Morin2022, Renger2022}.
Phenomenological equilibrium models, which account for polymer entropy, surface
tension, and droplet free energy, provide a framework for predicting phase
transitions. However, they do not explain how clusters of proteins and nucleic
acids dynamically reorganize to reach equilibrium. Addressing this question is
essential for linking \textit{in vitro} observations to the complex,
non-equilibrium behavior of chromatin in living cells.

Pioneering work by De Gennes \cite{DeGennes1985} and others \cite{Klushin1998,
Halperin2000} explained how strong attractive interactions between segments of a
freely draining chain collapses the polymer in a multi-step process.
Later confirmed by simulations, this process begins as an initially extended
polymer reconfigures into a two-phase structure consisting of dense clusters
connected by stretched segments. In this long-lived ‘pearl-necklace’ phase, the
polymer ends are gradually drawn together as the system relaxes, decreasing the
number of clusters over time and increasing the fraction of polymer
in the dense phase \cite{Halperin2000}. The reduction in cluster number is
governed by two competing mechanisms: direct coalescence of spatially adjacent
clusters and a ‘curious’ Ostwald ripening process, wherein polymer is pulled
from one cluster into another \cite{Klushin1998}. The relative contributions of
these mechanisms and the microscopic determinants that drive polymer
reorganization remain open questions, especially in biological contexts.
    
When a polymer is constrained at both ends— as sections of chromatin may be when
tethered to the nuclear periphery by lamin fibers or DNA manipulated in optical
trap experiments—the reduction in cluster number slows, and the total amount of
polymer in the clustered phase becomes limited
\cite{Chauhan2022,Halperin2000,Mahajan2022}. Previous studies on constrained
polymer systems found that dense clusters diffuse along the polymer backbone,
facilitating the relaxation towards a monolithic cluster
\cite{Aranson2003,Singh2024,Takaki2024,Bunin2015}. However, these studies often
overlooked biologically relevant attributes of DNA-binding proteins, such as
limited binding valency between proteins or the collective viscoelastic
properties that emerge from protein-protein and protein-polymer association.
\cite{Strickfaden2020,Cao2019,Michieletto2022,Strom2024,Muzzopappa2021}.
Moreover, the time integration methods used in early computational studies are
known to overestimate diffusive behavior, leading to conclusions that neglect
the ripening mechanisms mentioned above. While later studies have addressed some
of these limitations, they have not directly reported on the basic determinants
and dynamics of homopolymer relaxation \cite{Khanna2019, Owen2023, Hult2017,
Walker2019a}. In order to accurately model biofilaments, it is essential to
incorporate physical principles intrinsic to DNA binding proteins and their
interactions. 

To model large numbers (hundreds to tens of thousands) of mesoscale biological
structures (e.g., nucleosomes, IDRs, HP1$\alpha$) in a computationally tractable
way, we turn to coarse-graining methods. One approach to capturing transient
interactions is the sticker-spacer model, which simplifies amino acids in IDRs
into different species of interacting spheres, greatly improving computational
efficiency compared to atomic-level molecular dynamic simulations. However, the
size disparity between histones and amino acids remains a challenge, preventing
these simulations \cite{Martin2020, Li2023} from reaching biologically relevant
timescales (minutes to hours). To overcome this limitation, we couple
coarse-grained polymer dynamics with a kinetic Monte Carlo scheme that models
protein-mediated crosslinks as transient harmonic springs between DNA segments.
Binding and unbinding events obey detailed balance, ensuring that protein
association does not artificially alter polymer relaxation dynamics
\cite{Khanna2019, Lamson2021}. We employ a novel semi-implicit collision
resolution algorithm to accurately resolve excluded volume interactions between
polymer segments. This prevents unphysical behavior such as complete system
collapse while still allowing the system to reach equilibrium \cite{Yan2022}.
Unlike conventional hard-core potential methods, our approach avoids numerical
stiffness and remains computationally efficient and accurate. 

Using our simulations, we demonstrate how the interplay between polymer
diffusion, protein-mediated crosslinking, and spatial constraints governs the
macroscopic condensation dynamics. We identify the collapse pathways of flexible
biopolymer driven by transiently crosslinking proteins in both \textit{in vitro}
and \textit{in vivo} systems. Building on these insights, we develop a minimal
model that predicts and explains the long-time behavior of collapsing polymers.
Combined with numerical simulations, this model unifies previous theoretical
frameworks for synthetic constrained polymers and the reorganization of
biopolymers under confined conditions. One key advance is the ability to
directly link the microscopic properties of DNA-binding proteins to the
macroscopic and phenomenological parameters commonly used in DNA
dynamics research \cite{Quail2021,Renger2022}.

\section{Brownian Dynamics Model (\textit{aLENS})}%
\label{sec:model}
\begin{figure}[h]
  \centering
  \adjustbox{trim={0.0\width} {0.0\height} {0.0\width} {0.0\height},clip}
  {\includegraphics[width=3.25in, angle=0]{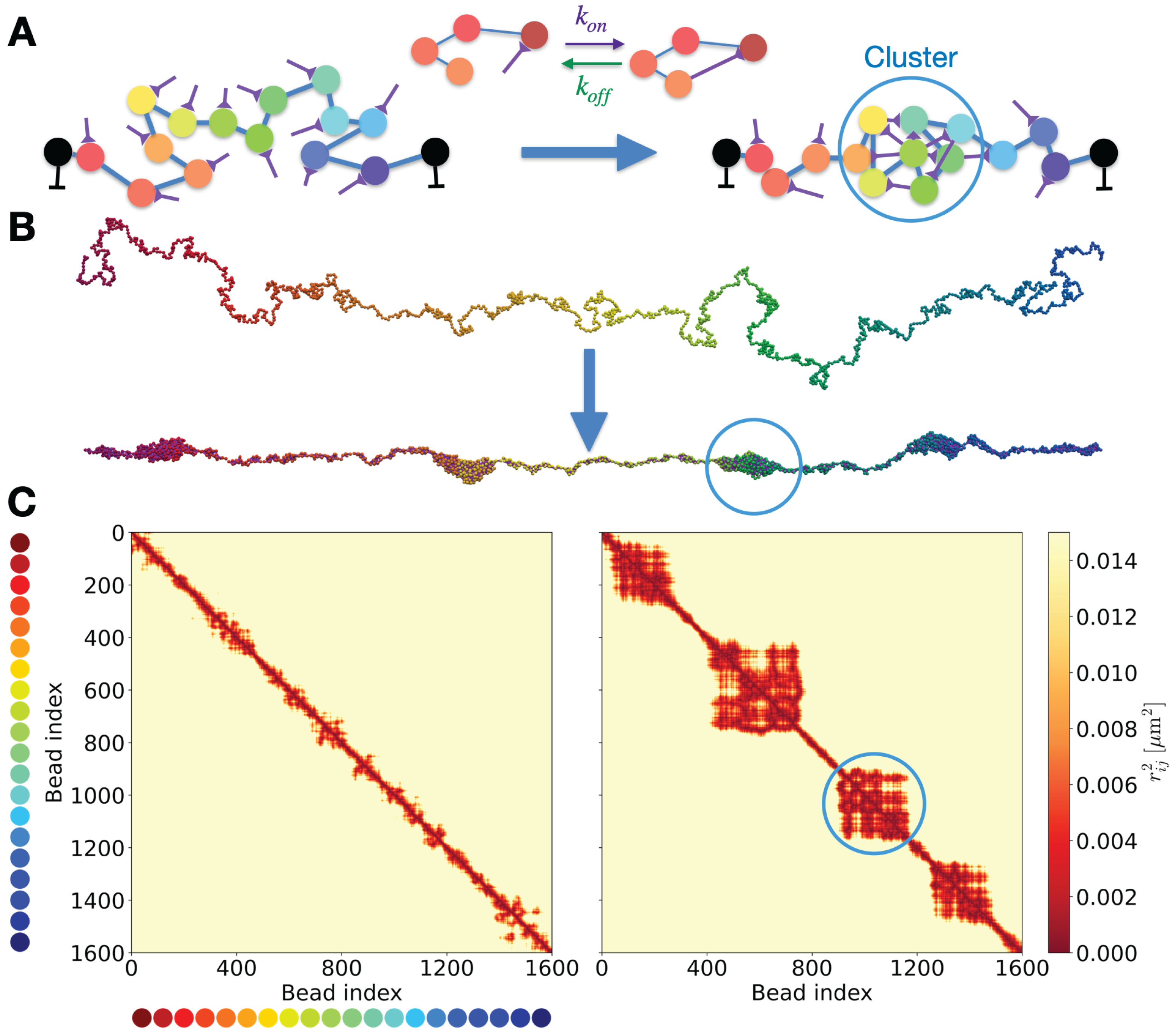} }
  \caption{ (A) Schematic representation of a polymer chain with sticky tails
  and fixed ends. The color of the beads indicates their index, with black
  representing the two fixed ends. Sticky tails are depicted as purple bars
  attached to the beads. Under the influence of kinetic binding (indicated by
  purple and green arrows), sections of the chain form dense clusters (light
  blue circle). (B) Simulation snapshots at $t=0\, \mathrm{s}$ (top) and after
  undergoing condensation at $t=600\, \mathrm{s}$ (bottom). An example cluster
  is circled in light blue. (C) Bead distance maps from the snapshots shown in
  (B). The left (right) plot is calculated from bead positions at $t=0\,
  \mathrm{s}$ ($t=600\, \mathrm{s}$). The circled cluster in (B) corresponds to
  the region marked in light blue in the right plot.  
  }
  \label{figs:fig1}
\end{figure}
\FloatBarrier
To model constrained polymer collapse driven by crosslinking proteins, we consider a self-interacting, freely draining bead-spring chain. For a chain consisting of $N$ beads, the system configuration is represented by the vector
\begin{equation}
    \bmc{C} = [\bm{R}_1, \ldots, \bm{R}_N]^T \in \mathbb{R}^{3N},
\end{equation}
where $\bm{R}_i$ is the 3D position vector of the $i$-th bead. The system's motion is governed by the overdamped Langevin equation:
\begin{equation}
  \dot{\bmc{C}} = \bmc{U} = \bmc{M}\, (\bmc{F}_B + \bmc{F}_C + \bmc{F}_X ) + \bmc{U}_n,
\end{equation}
where $\bmc{U}$ is the the grand velocity vector, $\bmc{M}\in \mathbb{R}^{3N
\times 3N}$ is the mobility matrix, and $\bmc{F}_{B,\,C,\,X} \in
\mathbb{R}^{3N}$ are forces arising from harmonic springs connecting beads in
the polymer chain, collision forces, and forces due to crosslinking,
respectively. The term $\bmc{U}_n$ accounts for thermal
fluctuations and satisfies the fluctuation-dissipation theorem by being drawn from a
Gaussian distribution such that
\begin{equation}
    \langle \bmc{U}_n \rangle=\bm{0} \qquad \langle \bmc{U}_n(t) \, \bmc{U}_n(t')^T \rangle = 2\,k_b\, T  \,  \bmc{M} \, \delta(t-t'),
\end{equation}
with the Boltzmann constant $k_b$ and temperature $T$. Hydrodynamic interactions
between beads are neglected. As a result, the mobility matrix for a freely
draining chain with spherical beads is diagonal, with elements given by the
Stokes-Einstein relation $\bmc{M}=(3\pi\eta b)^{-1} \bm{I}^{3N} $, where
$\bm{I}^{3N}$ is the $3N\times 3N$ identity matrix, $\eta$ is the
viscosity of surrounding fluid, and $b$ the diameter of the beads. To replicate
the experimental conditions involving constrained or tethered chains, the first
and last beads of the chain are fixed in space (Fig.~\ref{figs:fig1}). For these
fixed beads, the associated mobilities and noise terms are set to zero. 

The positions of beads are integrated through a
linearized Euler time-stepping scheme \cite{Yan2022} 
\begin{equation}
   \bmc{C}^{k+1} = \bmc{C}^k +  \bmc{U}^k \Delta t 
\end{equation}
However, forces arising from harmonic bonds and those due to hard collision
potentials introduce numerical stiffness to the problem, restricting the
timestep size when treated explicitly. Therefore, we calculate the bead
velocities $\bmc{U}^{k*}$ by solving a constraint optimization problem. This
approach improves numerical stability and allows us to achieve biologically
relevant timescales in simulations. We enforce hard steric interactions, which
prevents bead overlaps and permits chain entanglement, through a complementarity
constraint. For bead pairs $i,j$. separated by distance $r_{ij}=|R_i - R_j|$ we
ensure that the inequalities
\begin{equation}
  \begin{aligned}
    \text{No contact: } & r_{ij}-b\geq 0, \quad f_{C,ij}=0, \\
    \text{Contact: }    & r_{ij}-b=0, \quad f_{C,ij}\geq0,
  \end{aligned}
  \label{flip_flop}
\end{equation}
are satisfied at every timestep by solving for $\bmc{U}$. $\bm{f}_{C} \in
\mathbb{R}^{M}$ is the vector of constraint force magnitudes with dimensionality
$M$ equal to the number of pairs within a search radius set for computational
efficiency. The directed constraint forces are calculated by constructing a
sparse directed matrix $\bmc{D}(\bmc{C})\in \mathbb{R}^{3N \times M}$ such that
$\bmc{F}_C = \bmc{D}\bm{f}_{C}$. 

All bonds are taken to be harmonic with a given rest length resulting in a force
magnitude matrix 
\begin{equation}
  \bm{f}_B = -\bmc{K}_{B}(\bm{r}_B-b\bm{1}^{N-1})\quad \text{and} \quad  \bm{f}_X = - \bmc{K}_{X}(t)(\bm{r}_X-l_X\bm{1}^{N_X}),
\end{equation}
where the polymer stiffness matrix $\bmc{K}_{B}$ and crosslinking matrix
$\bmc{K}_{X}$ are diagonal with entries equal to the spring constants
$\kappa_{DNA}$ and $\kappa_X$, respectively (Table \ref{tab:parameters}). Here
$\bm{1}^N=[1,\, \ldots,\, 1]^T\in \mathbb{R}^N$, and $\bm{r}_{B}=[r_{0\,1},\, \ldots, \,
r_{i\,i+1}, \, \ldots, \, r_{N-2\,N-1}]^T\in \mathbb{R}^{N-1}$ and
$\bm{r}_{X}\in \mathbb{R}^{N_X}$ contain the distances between pairs of
permanently and transiently bonded beads, respectively. To form a polymer chain,
$\bm{r}_{B}$ is defined so beads are connected to neighbors in index space with
rest lengths equal to the bead diameter. Crosslinking bonds have a fixed rest
length $\ell_X$. The transient nature of the crosslinking bonds means that the
size of $\bmc{K}_X$ changes, $N_X=N_X(t)$. The direction of these forces is
assigned similarly to collision forces. At each timestep, the forces $\bm{f}_C$, $\bm{f}_B$, and $\bm{f}_X$ are computed by formulating the discretized equations of motion as a quadratic programming problem \cite{Yan2019, Yan2022}. Both spring and collision forces are resolved at every timestep using a fast global optimization algorithm.

The polymer chains are subject to transient crosslinking or \textit{sticky tails}. These sticky tails model biologically reactive components such as intrinsically disordered regions (IDRs) of DNA/chromatin binding proteins, linker histone H1, or histone tails \cite{Pepenella2014,Frank2020, Papamokos2012, Martin2020, Renger2022, Quail2021}. These tails are simplified to allow simulation of systems with $10^3 - 10^4$ beads over timescales ranging from minutes to hours. The tails can be viewed as harmonic springs with one end permanently bound to a base bead (the \textit{base end}) and the other end capable of binding and unbinding to neighboring beads (the \textit{transient end}) When only the base end is bound, the tail exerts no force on the polymer. However, the tail's transient end binds to other beads probabilistically according to an Arrhenius-like law:
\begin{equation}
    \label{eq:kon}
  k_{on}(r_{ij}) = k_o \exp\left(-\frac{\Delta G(r_{ij})}{k_BT}\right),
\end{equation}
where $k_o$ sets the timescale of the reaction. This treatment implicitly satisfies detailed balance if the transient end unbinds at the rate $k_{off}=k_o$ and the difference in free energy takes the form
\begin{equation}
\label{eq:deltaG}
  \Delta G(r_{ij}) = \epsilon_1 + \frac{\kappa_x}{2}\left(r_{ij} - l_X\right)^2.  
\end{equation}
The first term, $\epsilon_1$, in equation \eqref{eq:deltaG} accounts for the
binding energy and is associated with the binding equilibrium constant $K_e$ by
$\epsilon_1 = \ln\left(\frac{k_{on}(0)}{k_{off}}\right) =\ln(K_e)$. The
second term is the enthalpic contribution from the spring's potential energy.
For simplicity, force-dependent unbinding is not considered in this model.
Although there is no explicit limit on the number of transient ends that can
attach to a bead, steric interactions limit the local density of neighboring
beads. Given the exponentially decaying form of the binding rate in equation
\eqref{eq:kon}, the effective maximum number of attachments is set by the tails'
spring constant, rest length, temperature, bead diameter, and number of tails
per bead $n_x$. According to this model, for weak springs or long rest lengths,
the tails may bind through beads in dense clusters. To avoid unphysical
behavior, we have selected values for tail lengths and spring constants that are
consistent with biological structures (Appendix
\ref{sub:parameter_calculations}). 

\section{Results}%
\label{sec:results}

Here, we investigate a constrained polymer chain whose condensation is induced
by transient crosslinking proteins. The self-organization driven by transient
bonds accounts for the attractive interactions typical of intrinsically
disordered regions (IDRs) found in transcription factors and other DNA-binding
proteins \cite{Pepenella2014, Wang2018, Martin2020, Brangwynne2015, Renger2022,
Quail2021}. The chains' fixed ends simulate scenarios such as optical trap
experiments or nuclear environments, where condensing DNA regions are restricted
by tethering or frictional forces.

Experimental evidence suggests that both mechanical constraints and kinetic
interactions are important for protein-mediated biopolymer organization
\cite{Falzone2012, Dmitrieff2017, Bashirzadeh2021,Keenen2021}. We examine, in
particular, how the association constant $K_e$ of transient bonds influences the
nucleation, growth, and disappearance of condensed regions, or \textit{clusters},
when ends are held at a fixed distance $L_{sep}$. Each simulated chain consists
of 1600 beads with a diameter of $b = 0.01\,\mu\mathrm{m}$, resulting in a total
contour length of $L_{\mathrm{tot}} = 16\,\mu\mathrm{m}$. Clusters are
identified using the DBSCAN algorithm (see Appendix
\ref{sub:appendix:cluster_def_tracking:DBSCAN}), enabling us to quantify dynamic
properties such as the number $N_c(t)$ and  size distribution of clusters
$P(\ell_i,t)$. We conduct a preliminary investigation into the effects of
varying tail length, spring constant, number of tails per chain segment, and
overall chain length. Given the computational expense of a comprehensive
analysis across all parameters, we focused on a baseline parameter set that
reflects behavior observed in previous \textit{in vitro} experiments (see Table
\ref{tab:parameters}) \cite{Quail2021,Renger2022}. We then characterize the
effect of binding affinity, which is an experimentally relevant variation. 

\subsection{Determinants of Polymer Clustering}
\label{sub:determinant_of_clustering}

\begin{figure}[h]
  \centering
  \adjustbox{trim={0.0\width} {0.0\height} {0.0\width} {0.0\height},clip}
  {\includegraphics[width=\columnwidth, angle=0]{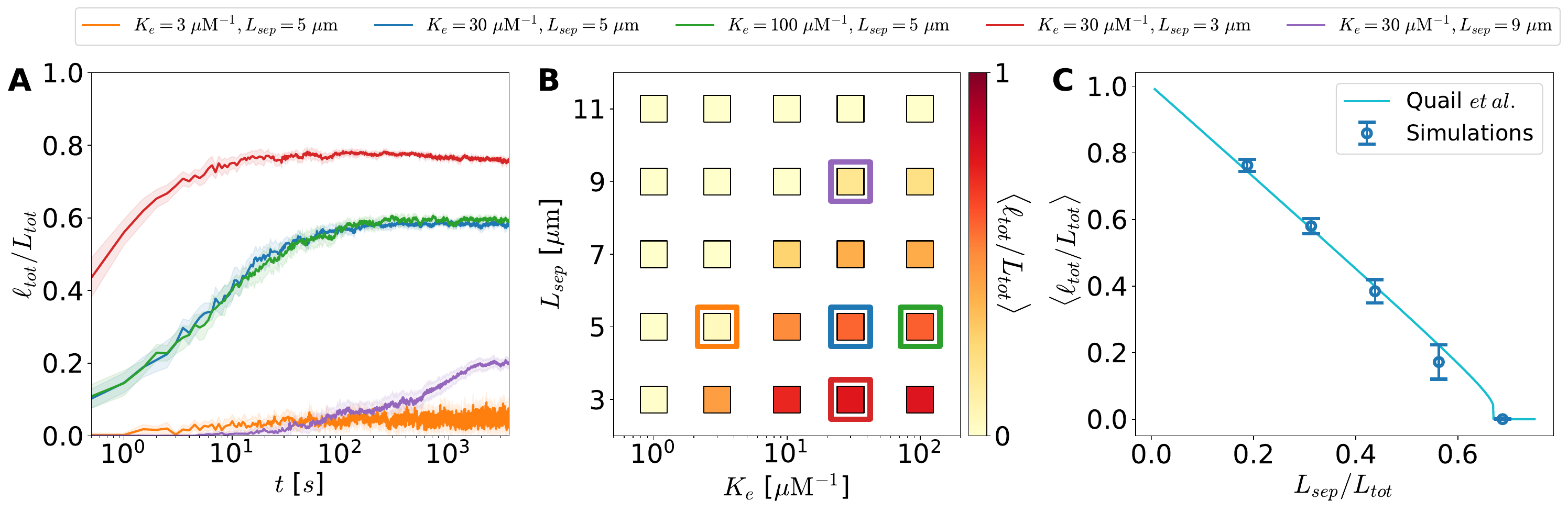} }
  \caption{ (A) Average values of total clustered chain length $\ell_{tot}$
  normalized by $L_{tot}$. Curves represent averages over 12 independent
  realizations. Curve colors correspond to different values of $K_e$ and
  $L_{sep}$, as indicated at the top of the figure. Shaded areas are $95\%$
  confidence intervals of the mean. (B) Equilibrium values of $\langle
  \ell_{tot}/L_{tot} \rangle$ for different values of $K_e$ and $L_{sep}$. The
  and the colored boxes around data points correspond to the curves in (A). (C)
  Comparison of the total clustered chain length from our $K_e=\SI{30}{\per\micro\molar}$ simulations (dark blue) to previous equilibrium theories with experimentally inferred constants
  (cyan) \cite{Renger2022, Quail2021}.}
  \label{figs:fig2}
\end{figure}

To ensure that the averaged results are independent of initial configurations,
the polymer chain is first set in a helical geometry with a fixed number of
turns (Appendix \ref{ssub:nucleating_cluster_simulations}). The chain is then
evolved under Brownian fluctuations (without crosslinking) for $T = 100$
seconds, or $T/\tau_R \approx 1$, where $\tau_R=\frac{\eta b^3N^2}{\pi k_B T}$
is the Rouse relaxation time. Twelve independent realizations of the system are
averaged to obtain the final results, along with confidence intervals as shown
in Fig.~\ref{figs:fig2}(A).

We first examine how the condensation dynamics depend on $K_e$ and $L_{sep}$.
Fig.~\ref{figs:fig2}(A) shows temporal courses of condensation, as measured by
$\ell_{tot}$, the total length of chain in clusters. It is only beyond a
critical value of $K_e*$ that clusters form, with $\ell_{tot}$ reaching a
plateau for a given $L_{sep}$.  Fig.~\ref{figs:fig2}(B) exhibits the phase
transition behavior of the polymer chain by showing the fraction of the chain in
clusters as a function of $K_e$ and $L_{sep}$. Similar phase transitions from
unclustered to clustered states are predicted by phenomenological models
and observed in experimental studies \cite{Renger2022, Quail2021}. From
Fig.~\ref{figs:fig2}(A,B), we determine the critical values of the microscopic
parameters, $K_e^* = \SI{1}{\per\micro\molar}$ and
$L_{sep}^*=\SI{11}{\micro\meter}$, that determine the clustered state of a
constrained chain at long times. Further details on how variations in $K_e$
affect cluster dynamics are discussed in Sec.~\ref{sub:cluster_evolution}.
Additionally, our results show that increasing the end-to-end separation
$L_{sep}$ slows clustering, as is evident from the comparison of the red, blue,
and purple curves in Fig.~\ref{figs:fig2}(A). 

Fig.~\ref{figs:fig2}(C) shows that the long-time behavior of $\ell_{tot}$ in
simulations with $K_e = \SI{30}{\per\micro\molar} \, \mu M^{-1}$ aligns well
with the equilibrium model and parameters proposed and measured by Quail
\textit{et al.} to describe \textit{in vitro} experiments \cite{Renger2022,
Quail2021}. Their model relied on fitted, phenomenological parameters, such as
surface tension and condensate free-energy density, to predict the likelihood of
DNA-protein co-condensation (Fig.~\ref{figs:fig2}(C) cyan). These equilibrium
theories describe the steady-state macroscopic properties of the system, such as
$\ell_{tot}$, but do not address the transient dynamics of clusters. Our
approach captures both the macroscopic properties of the system and the cluster
evolution. The agreement between our simulations and the equilibrium model
suggests a correspondence between the microscopic and macroscopic parameters
employed in both approaches. In the following sections, we explore how systems
of clustering chains progress toward their steady-state configurations.

\subsection{Spatiotemporal Dynamics of Cluster Evolution}
\label{sub:cluster_evolution}

\begin{figure}[h!]
  \centering
  \adjustbox{trim={0.0\width} {0.0\height} {0.0\width} {0.0\height},clip}
  {\includegraphics[width=3.25in, angle=0]{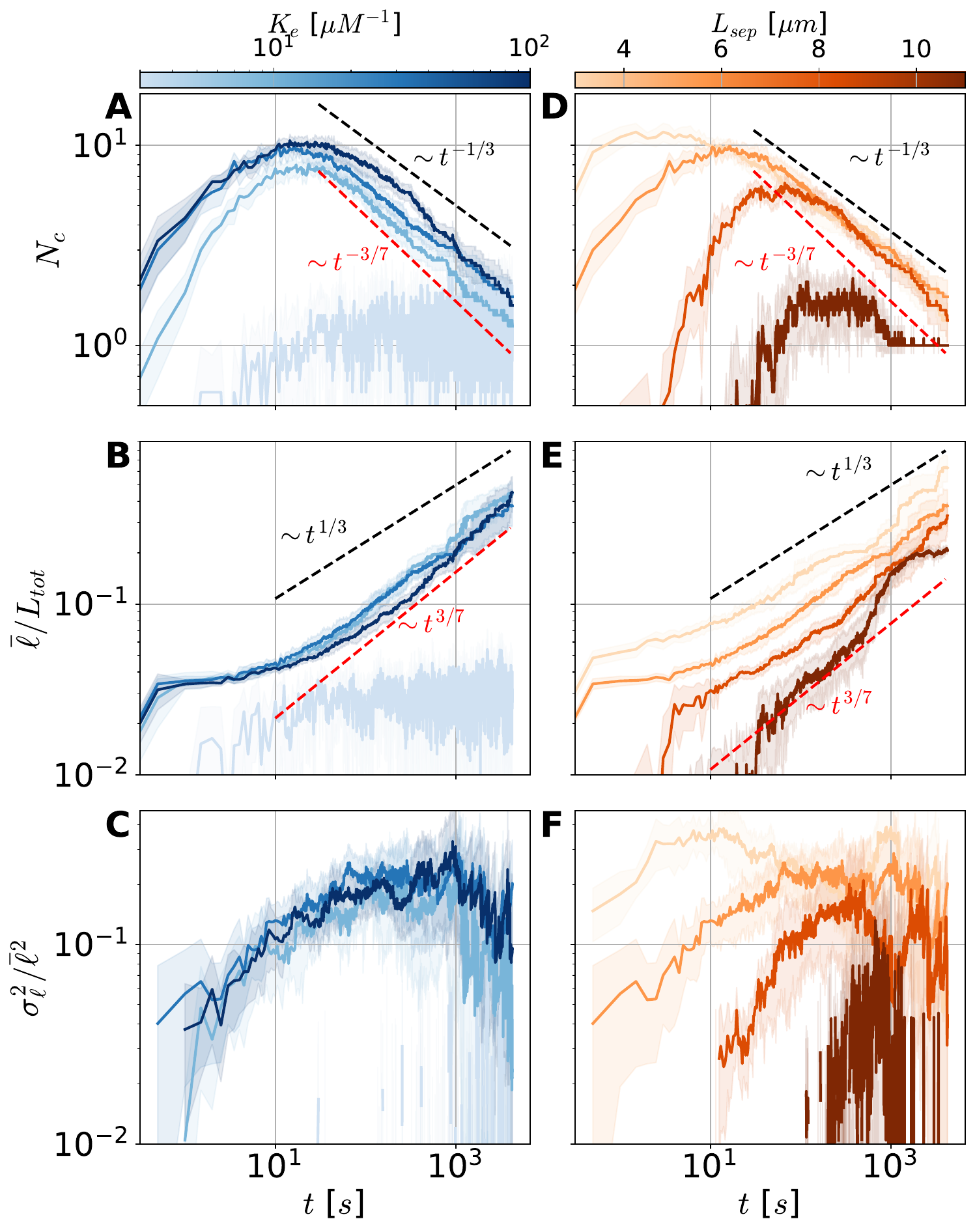} }
  \caption{(A,D) Number of clusters $N_c$, (B,E) average cluster size $\barell$, and
  (C,F) variance of clusters $\sigma_\ell^2$ scaled by $\barell^{-2}$ as functions of time. Curves are averaged over 12
  realizations using different random number generator seeds. Shaded areas are
  95\% confidence intervals of the mean. (A-C) Parameter sets with
  $L_{sep}=\SI{5}{\micro\meter}$ and varied binding affinities ($K_e = 3, 10,
  30, 100 \mu M^{-1}$).  (D-F) Parameter sets with $K_{e}=30 \mu M^{-1}$ and
  varied end separations ($L_{sep} = 3,5,7,9 \mu\text{m}$).}
  \label{figs:fig3}
\end{figure}

During simulations, multiple clusters rapidly form along the chain at short
times, eventually evolving toward a single cluster at longer timescales ($t \gg
\tau_R$) as shown in Fig.~\ref{figs:fig3}(A,D). This dynamic behavior is
consistent across all systems exhibiting clustering and has been previously
observed in both free and constrained, potential-based polymer simulations
\cite{Aranson2003, Singh2024,Lappala2013a,Takaki2024}. As we did for the
equilibrium cluster states in section \ref{sub:determinant_of_clustering}, we
now wish to characterize how variations in $K_e$ and $L_{sep}$ influence the
dynamics of individual clusters.

We first examine how the number of clusters evolves over time as a function of
tail binding affinity. As previously mentioned, we observe distinct clustering
regimes. At $K_e = 1 \, \mu M^{-1}$, no clustering occurs while at $K_e = 3 \,
\mu M^{-1}$, transient individual clusters with short lifespans form. For higher
binding affinities ($K_e \geq 10 \, \mu M^{-1}$), we observe that clusters
nucleate and grow (see Fig.~\ref{figs:fig2_supp_contact_kymo_mod} in Appendix
\ref{sub:appendix:cluster_def_tracking:temporal_track_history}). For $K_e \ge 10
\, \mu M^{-1}$, the maximum number of nucleated clusters increases only modestly
with higher binding affinity, as shown in Fig.~\ref{figs:fig3}(A).

Following the initial nucleation of clusters, we observe a universal decrease in
cluster number over time, as shown in Fig.~\ref{figs:fig3}(A,D). The number of
clusters appears to follow a power law $N_c \propto t^{-\mu}$ in this regime,
with an exponent in the range $1/3 \leq \mu \leq 3/7$. Previous simulations
using Lennard-Jones potentials by Aranson and Tsimring \cite{Aranson2003} showed
a somewhat faster decrease in the number of clusters following $N_c \propto
t^{-1/2}$. This behavior was attributed to the merging of clusters driven by
diffusion along the chain backbone. However, the Aranson and Tsimring noted that
cluster diffusion was likely overestimated due to the use of the Nose-Hoover
thermostat, which could bias the rate of cluster disappearance. This discrepancy
in the scaling law for $N_c$ between previous works and our simulations suggests
the presence of additional mechanisms influencing the growth and disappearance
of clusters. Interestingly, at longer times -- when the total chain in clusters
has plateaued $>$\SI{10}{\sec})-- the cluster variance appears to scale with the
average cluster size squared Fig.~\ref{figs:fig3}(C,F).  This information proves
helpful in deriving a theory to describe the evolution of cluster distributions.

Using free energy arguments to construct a minimal model, we explain the
clustering dynamics observed in our simulations (see Appendix
\ref{sec:minimal_model} for details). We assume the rate at which chain segments
enter clusters is determined by the balance between the tension exerted by
clusters on unclustered chain sections (strands) and the viscous friction on the
strands from the solvent. The force generated by cluster $i$ on its adjacent
strands is derived from the free energy of a cluster as 
\begin{equation}
    \label{eq:cluster_force}
    f_i = -\alpha + \gamma \ell_i^{-\frac{1}{3}},
\end{equation}
where $\alpha$ represents the cluster's bulk energy and  $\gamma$ denotes the
effective surface tension. Consequently, we determine that the growth rate of
the average cluster size follows $\bar{\ell} \propto t^{\mu}$, where the
exponent $\mu$ depends on the primary mechanism driving cluster evolution.
Specifically, $\mu = 1/3$ corresponds to a merging-dominated regime, while $\mu
= 3/7$ indicates a ripening-dominated regime, where ripening refers to the
growth of some clusters at the expense of others shrinking and dissolving. The
characteristic timescales for these two processes can be estimated as follows:
\begin{equation}
    \tau_{merge} \cong \bar{s}^2/\bar{D}_c, \qquad \tau_{ripe}\cong {\eta
    \,\bars\, \barell^{4/3}}/{\gamma}, \label{eq:tau_merge_dissolve}
\end{equation}
where $\bar{D}_c \cong k_B\,T/\eta \,\bar{\ell}$ is the average diffusion
constant of clusters and $\bar{s}$ is the average length of the strands
connecting clusters. Initially, when clusters are small and positioned closely
along the chain, merging plays a dominant role. As the average strand
length between clusters increases, the merging timescale,
$\tau_{merge}$—governed by one-dimensional diffusion along the backbone—grows
and eventually increases beyond the timescale for ripening. Thus,
ripening becomes the dominant mechanism, as $\tau_{ripe} < \tau_{merge}$.


Extending our analysis, we examine the average length of chain in clusters and
in connecting strands, denoted by $\barell = \ell_{tot} / N_c$ and $\bars = s_{tot}
/ (N_c+1) \approx s_{tot} / N_c$, respectively. These definitions imply a
relationship between the average cluster size and the strand length, expressed
as $\bars \approx a \, \barell$, where $a = s_{tot} / \ell_{tot}$. The total
polymer length $L_{tot}$ can be separated into the contributions from
clusters and strands, yielding $L_{tot} = \ell_{tot} + s_{tot}$.
Moreover, as illustrated in Fig.~\ref{figs:fig2}(C), we observe that the total
length of clustered chains decreases nearly linearly with the end-to-end
separation, following the relationship $\ell_{tot} \approx L_{tot} - c \,
L_{sep}$, where $c$ is derived from Fig.~\ref{figs:fig2}(C) and is a function of
$K_e$. From this, we infer that $s_{tot} \approx c \, L_{sep}$, leading to:
\begin{equation}
    a = \frac{s_{tot}}{\ell_{tot}} \approx \left({\frac{L_{tot}}{c\,L_{sep}}-1}\right)^{-1},
\end{equation}
where $a$ is a function of $L_{tot}, K_e$ and $L_{sep}$. From this relation, we
may convert the timescales of equation \eqref{eq:tau_merge_dissolve} into
functions of only $\bars$, $\barell$, or $N_c$, allowing for analytic comparison
between timescales. For example, the relation highlights the
difference between the ripening mechanism of cluster growth and that
of typical Oswald ripening. By substituting $\barell$ for $\bars$ in equation
\eqref{eq:tau_merge_dissolve}, one finds $\barell \sim t^{3/7}$ through ripening
alone. Assuming that clusters are roughly spherical, this scaling implies an
average radius of clusters to scale as  $\bar{R}_c \sim t^{1/7}$ as opposed to
the classical Oswald ripening scaling of $\bar{R}_{Os} \sim t^{1/3}$.

\begin{figure}[h!]
  \centering
  \adjustbox{trim={0.0\width} {0.0\height} {0.0\width} {0.0\height},clip}
  {\includegraphics[width=3.25in, angle=0]{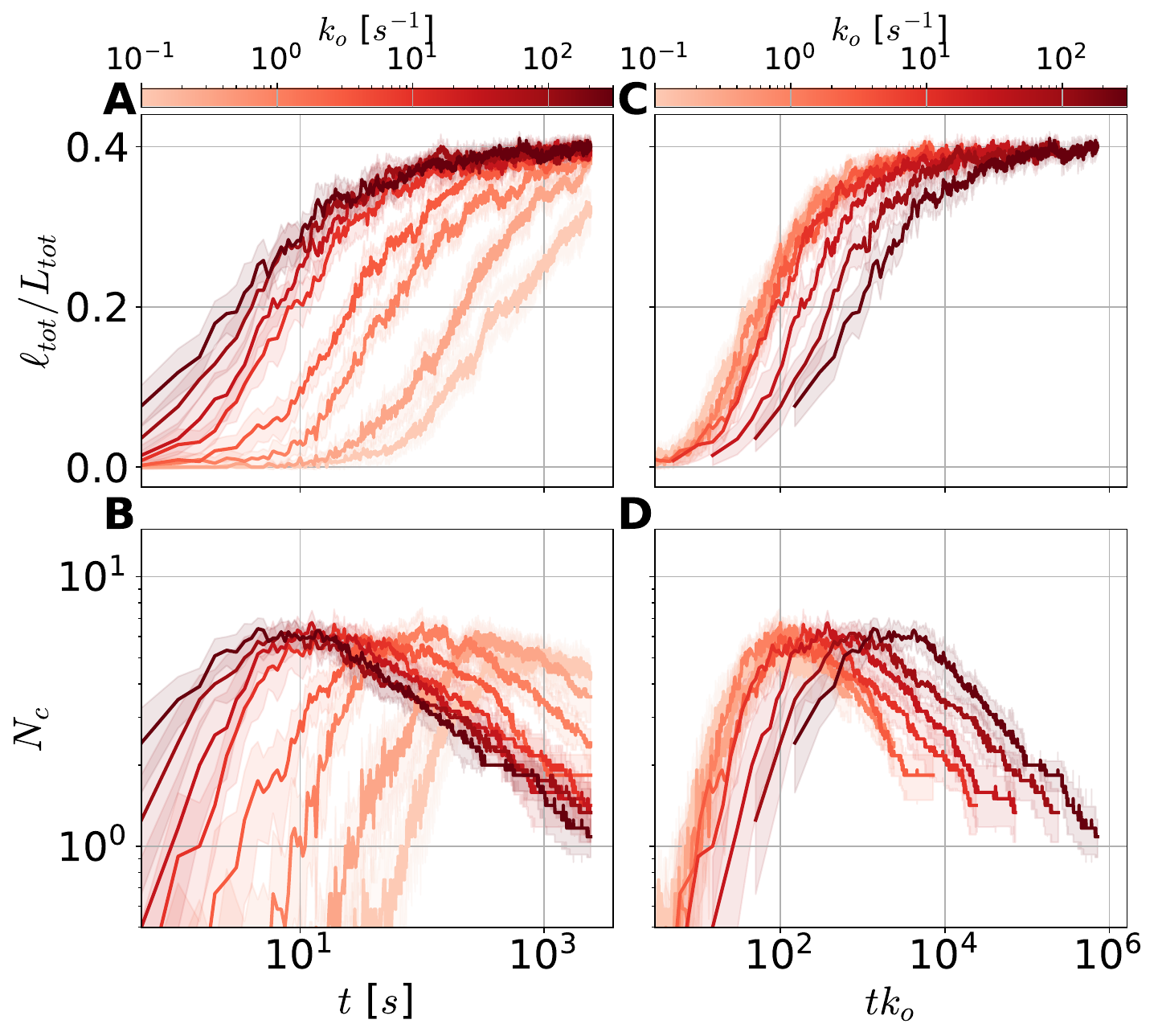} }
  \caption{(A,C) Total chain in clusters and (B,D) number of clusters as functions of time with $K_e = 30 \mu M^{-1}$,
  $L_{sep} = 5\mu m$, and varying crosslinker turnover rates
  ($k_o=.1,.3,1,3,10,30,100,300 s^{-1}$). Curves are averaged over 12
  independent realizations using different random number generator seeds. Shaded areas are
  95\% confidence intervals of the mean. In (A,B), time is in units of seconds
  whereas in (C,D), time is rescaled by the inverse of the crosslinker binding
  rate.
  }
  \label{figs:fig4}
\end{figure}

Next we investigate the effect of turnover rate, $k_o$, on system behavior
(Fig.~\ref{figs:fig4}). The turnover rate proportionally scales both the binding
and unbinding rates. Panels A, B in Fig.~\ref{figs:fig4} demonstrate that
increasing $k_o$ leads to higher rates of cluster formation, faster initial
cluster growth, and more rapid cluster disappearance at long times. However, the
maximum length of all clusters, the maximum number of clusters, and the scaling
behavior remain unchanged across all test cases. This confirms that these
quantities are primarily governed by $K_e$ and $L_{sep}$ rather than $k_o$.
Proportional changes in binding and unbinding rates alter the rate at which
macroscopic properties, such as the number of clusters, evolve but do not
necessarily affect the dominant mechanisms driving their evolution. Notably, for
$k_o \geq 10~{s}$, the measured quantities $\ell_{tot}/L_{tot}$ and $N_c$ show
similar behavior at times $>$\SI{10}{\sec} roughly collapsing onto a single
curve ( Fig.~\ref{figs:fig4}(C,D)). 

Conversely, when time is rescaled with $k_o$, simulations with $k_o < 10~s
$ also partially collapse onto a single curve, as shown in
Fig.~\ref{figs:fig4}(C,D). This suggests a crossover in the relevant timescales.
For slow crosslinking kinetics, cluster formation is reaction-limited. However,
as crosslinking becomes faster, the system transitions to a regime limited by
another mechanism. While several timescales are present in the system, the most
likely limiting factor is the viscous drag of the chains. Notably, the Rouse
time is unlikely to play a significant role in this crossover since the entire
chain does not need to rearrange for individual clusters to evolve. Instead, the
dynamics are governed by the motion of individual clusters and unclustered
segments. The drag coefficient for clusters or segments scales proportionally
with the number of beads they contain. Using the Stokes-Einstein relation, we
estimate the diffusion timescale over a distance equal to the bead diameter as
$\tau_d =3\pi \eta \barell b^2/k_B\,T$. For cluster or segment sizes on the
order of $100$ beads, this timescale is approximately $0.1~{s}$, which is
consistent with the critical turnover rate of $k_o \sim 10~{s}^{-1}$.

\subsection{Cluster Lineage and Modes of Growth}

While a minimal model, based on free energy arguments, successfully describes
the growth dynamics of clusters, it provides only partial insight into the
specific mechanisms governing cluster evolution and their relative significance.
To understand aspects such as cluster lifetimes, size fluctuations, and movement
in both positional and chain index space requires us to track individual
clusters throughout their evolution. Hence, we developed a tracking algorithm,
detailed in section \ref{sub:appendix:cluster_def_tracking:DBSCAN}, that
reliably recognizes clusters over time. This approach facilitates a more
detailed investigation of clusters' dynamic behavior.

\begin{figure}[h!]
  \centering
  \adjustbox{trim={0.0\width} {0.0\height} {0.0\width} {0.0\height},clip}
  {\includegraphics[width=3.25in, angle=0]{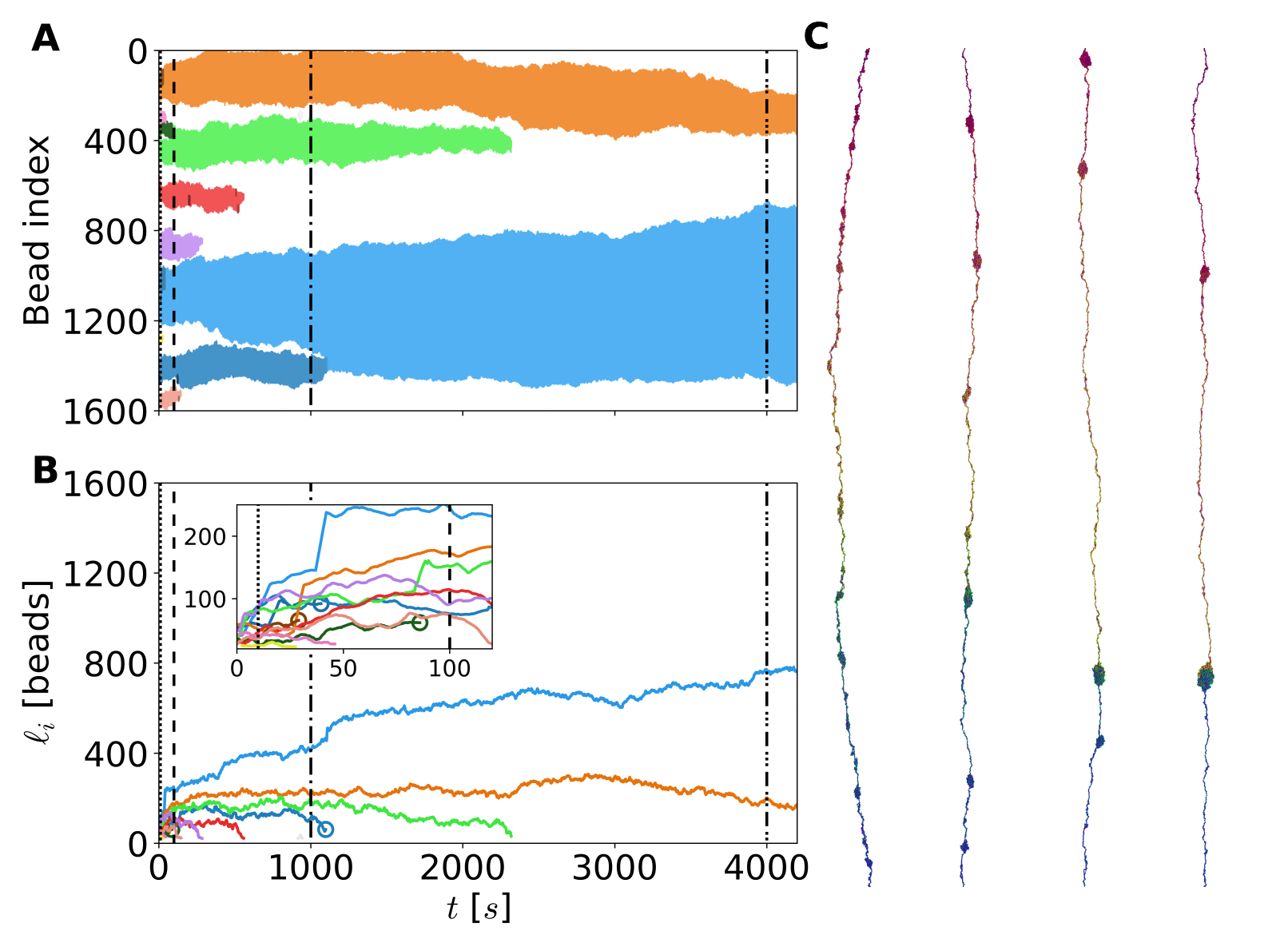} }
  \caption{Visualization of cluster dynamics in a single simulation. (A)
  Clusters shown in bead index space over time. Filled regions of the same color
  indicate bead indices belonging to the same cluster, with darker shades
  representing earlier time points in the cluster's genealogy. (B) Cluster sizes
  from (A) as a function of time, with merging events marked by circles. (Inset)
  Zoomed-in view of the first 100 seconds of (B). (C) 3D snapshot of the system
  at the time indicated by the dotted lines in (A) and (B).
  }
  \label{figs:fig5}
\end{figure}

Fig.~\ref{figs:fig5}(A) visualizes the evolution of clusters in the  index space
over time, showing the lineage of individual clusters. The corresponding cluster
sizes, shown in Fig.~\ref{figs:fig5}(B), track their nucleation, growth, and
disappearance over time. Using our algorithm, we determine whether a cluster
disappears due to merging with another cluster or by gradually dissolving as
beads dissociate. As is evident from Fig.\ref{figs:fig5}(A), a number of
clusters initially nucleate randomly along the backbone. Many of them rapidly
merge, forming larger clusters. At later times, $t > 100~s$, clusters not only
migrate through positional space but also translocate along the polymer backbone
as beads associate or dissociate (see
Fig.~\ref{figs:fig5_supp_tracking_multiple} in Appendix
\ref{sub:appendix:cluster_def_tracking:temporal_track_history}). 
At these later times, we also observe ripening phenomena as cluster sizes change
relatively slowly with larger clusters growing while the smaller clusters shrink
and dissolve. The larger the difference between clusters, the more likely and
more quickly the smaller cluster dissolves. Having looked at a single
realization, now we discuss the statistical behavior of clusters.

\begin{figure}[h!]
  \centering
  \adjustbox{trim={0.0\width} {0.0\height} {0.0\width} {0.0\height},clip}
  {\includegraphics[width=3.25in, angle=0]{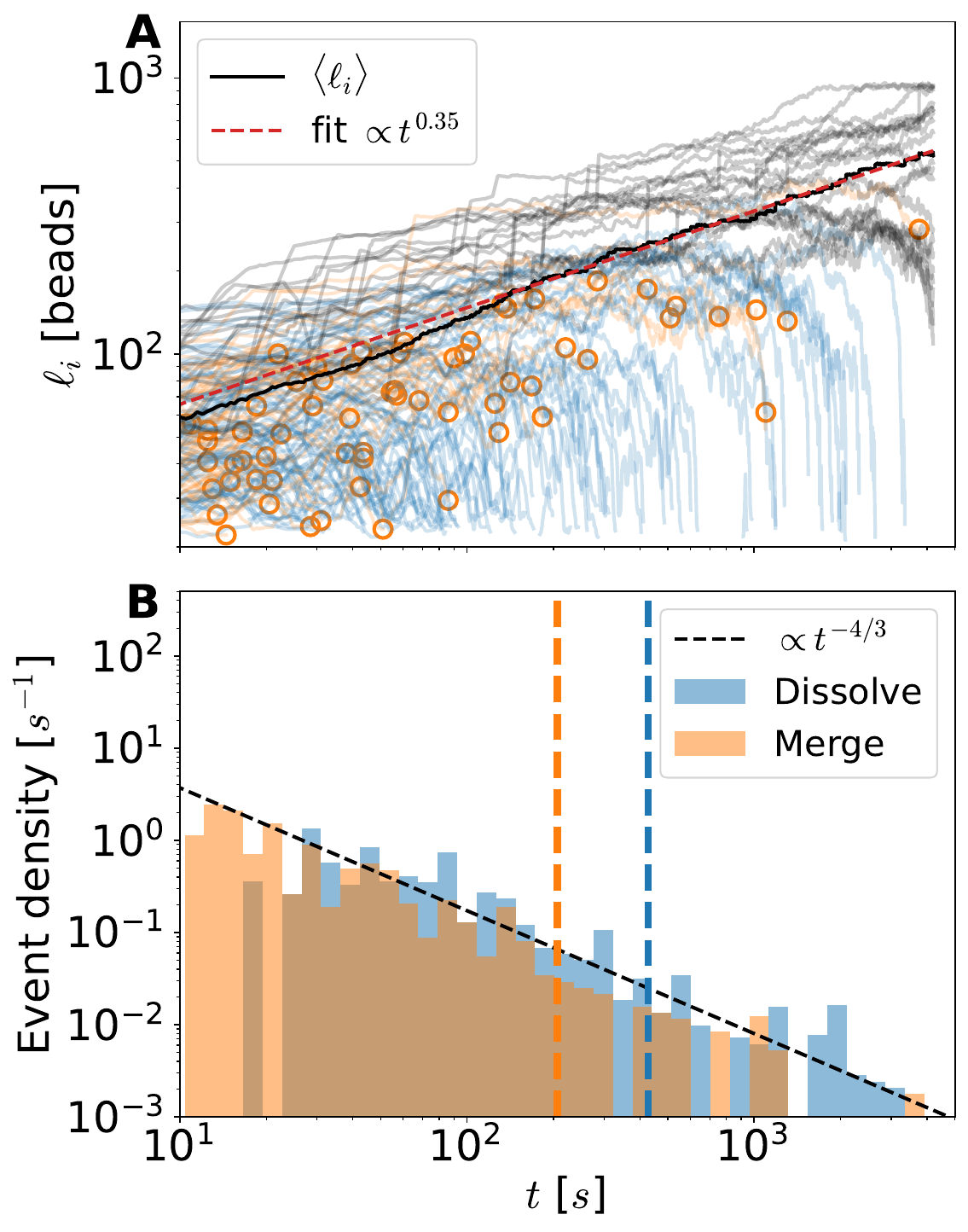} }
  \caption{ (A) Size traces of all clusters plotted as a function of time for
  systems with $L_{sep}=5\mu m$, $K_e = 30\mu M^{-1}$, and $k_o=3 s^{-1}$. Trace
  color indicates cluster fate: blue for dissolving, orange for merging, and
  gray for clusters that persist until the end of the simulation. Orange circles
  mark merging events. The average cluster size is plotted as a black line,
  fitted to a power law $\AVG{\ell_i} = at^b$ with $a = 29.37$, $b =.35$ and
  $r^2=.9905$ (red dashed line). (B) Histogram of dissolving (blue) and merging
  (orange) event times from (A). The mean merging time $\AVG{t_M}=206\;s$
  (dashed orange line) and  mean dissolving time $\AVG{t_D}=428\;s$ (dashed blue
  line) are shown along with the predicted scaling behavior (dashed black line).
  Proportion of events: dissolving $p_D= 0.59$ and merging $p_M=.41$.
  }
  \label{figs:fig6}
\end{figure}

The average cluster size, shown in Fig.~\ref{figs:fig6}(A), follows a scaling
law, $\langle \ell_i\rangle \propto t^{0.35}$, consistent with the
theoretical bounds discussed in Section \ref{sub:cluster_evolution}. Notably,
this scaling behavior differs from that found in previous studies, where cluster
growth was primarily driven by merging \cite{Aranson2003}. This led to an
average cluster length scaling as $\langle \ell_i\rangle\propto t^{1/3}$, though
it was acknowledged that the use of the Nose-Hoover integration method resulted
in artificially high cluster diffusion \cite{Aranson2003}. The authors of
\cite{Aranson2003} argued that in a minimal model, incorporating realistic
local drag would alter the cluster size scaling to $\langle \ell_i\rangle\propto
t^{3/7}$. Our work provides a physical foundation for this scaling law,
associating it with a regime where ripening is the dominant growth
mechanism.
 
Fig.~\ref{figs:fig6}(B) compares the histograms of merging and dissolving
events over time. In the first 20 seconds of the simulation, very few
dissolving events occur. During this period, the small clusters would only
experience weak tension 
from the excess chain between clusters, allowing clusters to grow unhindered.
Ripening begins in earnest only after the total chain length in clusters,
$\ell_{tot}$, plateaus and unclustered chain segments become stretched. As time
progresses, the average distance between clusters increases, making dissolving events
more frequent than merging.

Throughout most of the transient evolution of clusters in our simulations dissolving and merging events occur at comparable rates. At later times, both kinds of termination events become less frequent, though merging becomes less prevalent. There is little to no nucleation of new clusters at long times. From a free energy perspective, nucleating a new cluster would require overcoming a significant energy barrier, either by stretching the chain or reducing the size of existing clusters to accommodate a new cluster that does not immediately dissolve. Overall, ripening events events occur at a rate $1.43$ times that of merging events, though the mean dissolving time, $\langle t_D \rangle = 428~\mathrm{s}$, is longer than the mean merging time, $\langle t_M \rangle = 206~\mathrm{s}$.

\begin{figure}[h!]
  \centering
  \adjustbox{trim={0.0\width} {0.0\height} {0.0\width} {0.0\height},clip}
  {\includegraphics[width=\columnwidth,
  angle=0]{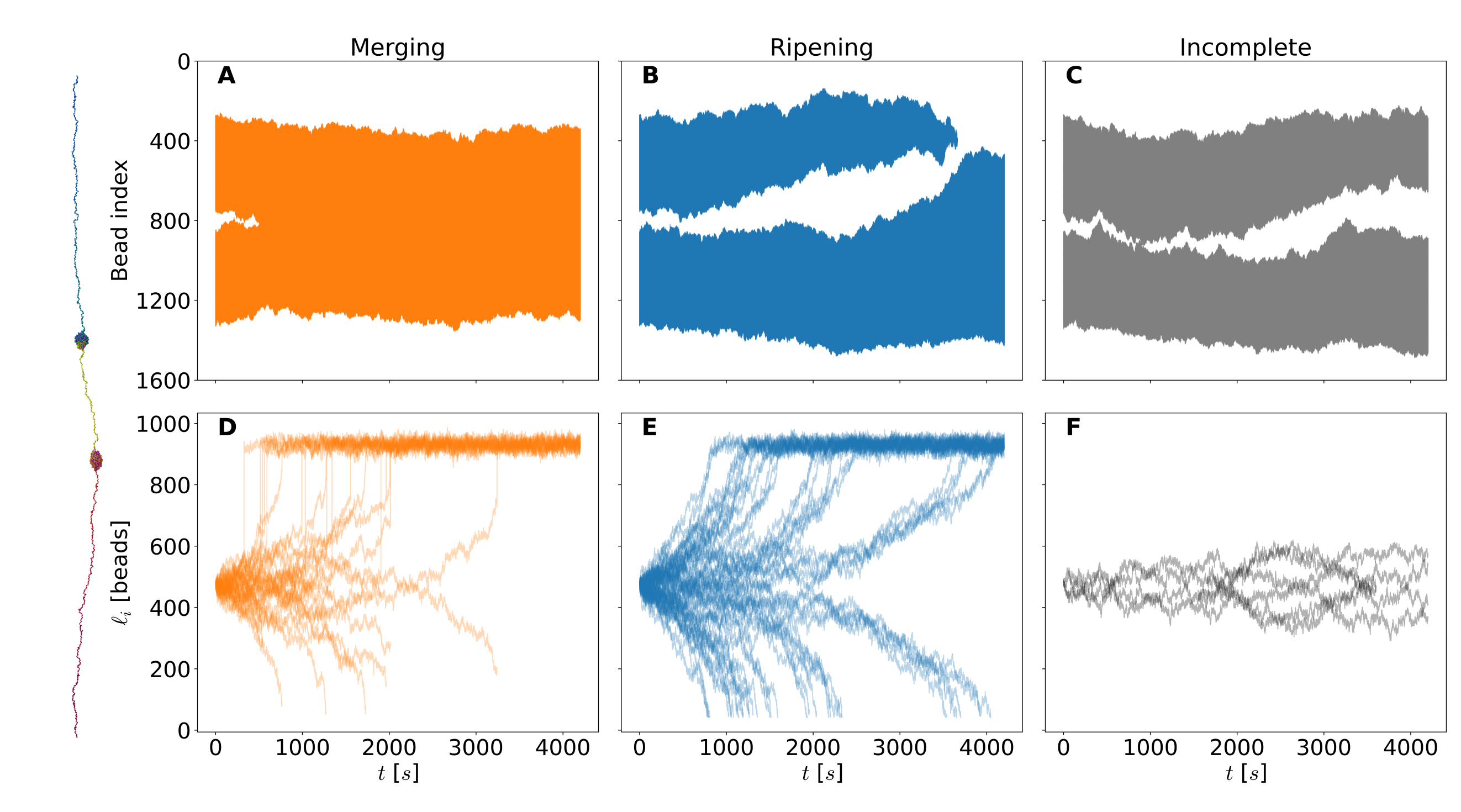} } \caption{Time
  evolution of equal-sized clusters separated by 100 beads, with $L_{sep}=5\mu
  m$, $K_e = 30\mu M^{-1}$, and $k_o=3 s^{-1}$, across 48 replicated
  simulations. (Far left) Example of the initial configuration in 3D. (A–C)
  Representative examples of clustered bead indices over time for (A) merging
  events, (B) ripening events, and (C) cases where no events occur. (D–F)
  Cluster size as a function of time for all replicates undergoing (D) merging
  events ($N_M =17$), (E) ripening (dissolving) events ($N_D=28$), and (F) no events
  ($N_{NE}=3$). }
  \label{figs:fig7}
\end{figure}

Since fewer clusters exist at later times, drawing convincing conclusions about
long-term cluster growth modes is challenging. To analyze late-time behavior in
more detail, we initialized simulations with two identically sized preformed
clusters separated by an unclustered chain segment of approximately 100 beads
($6\%$ of all beads), as shown in Fig.~\ref{figs:fig7}. The initial cluster
sizes were chosen so that the total chain length in clusters matched the
steady-state value found in Section \ref{sub:determinant_of_clustering}.

We observed that the segment length separating the clusters influenced both
dissolving and merging rates. Notably, when clusters were initialized with three
equal-length strands, no merging or dissolving events occurred within the first
$1200~\mathrm{s}$ (data not shown). However, reducing the strand length between
clusters to $100$ beads increased the frequency of events, consistent with our
estimates that predict $\tau_{merge} \sim s^2 \barell$ and $\tau_{ripe}\sim s
\barell^{4/3}$ from section~\ref{sub:cluster_evolution}. 

\begin{figure}[h!]
  \centering
  \adjustbox{trim={0.0\width} {0.0\height} {0.0\width} {0.0\height},clip}
  {\includegraphics[width=3.25in, angle=0]{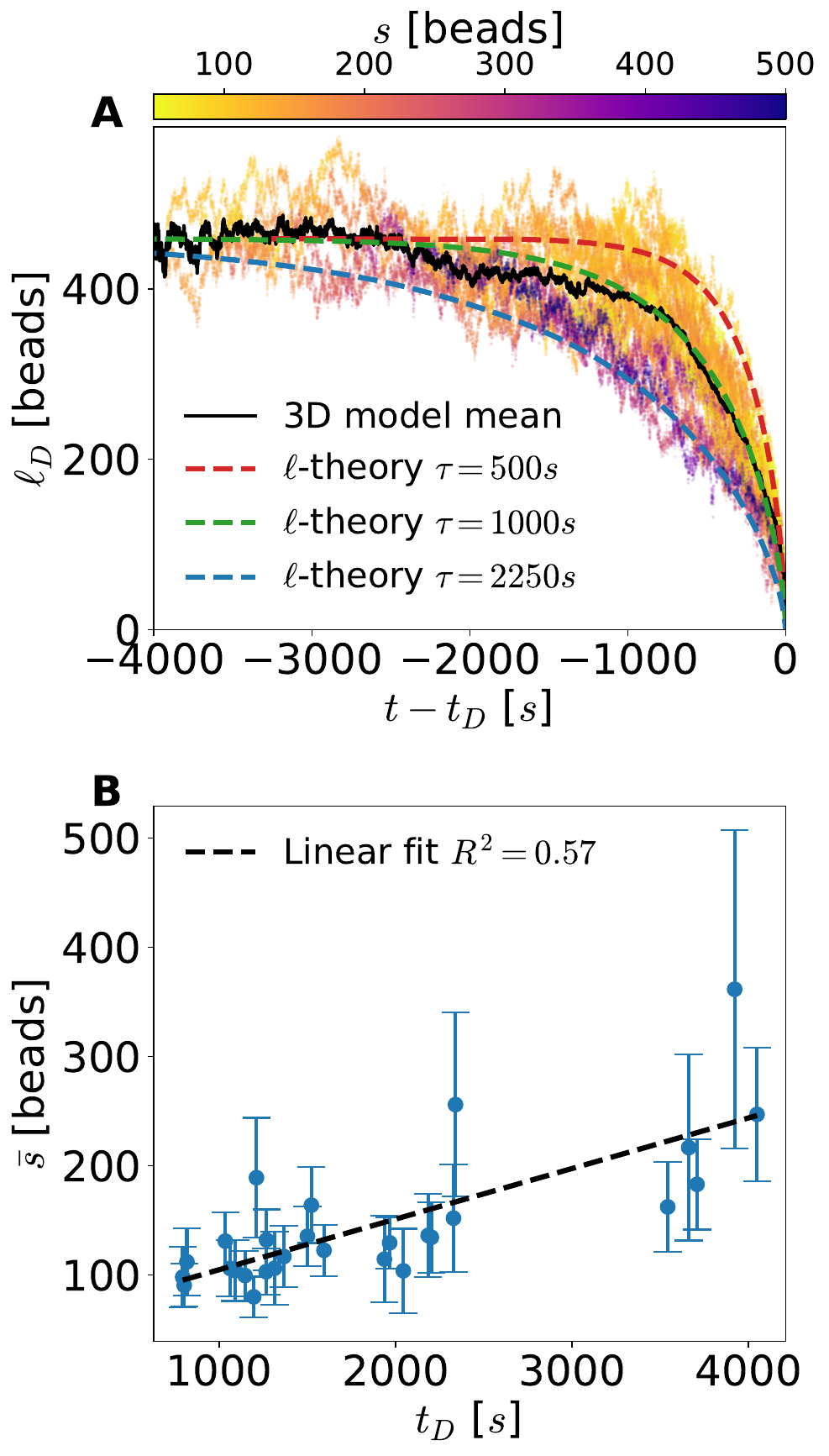} }
  \caption{Effect of strand length on the size evolution of dissolving clusters. (A) Synchronized cluster size traces for 28 dissolving clusters from Fig.~\ref{figs:fig7}(E). The size traces are temporally aligned such that the ripening time $t_d=0$. Line color indicates the instantaneous length of the strand separating the clusters. The black line represents the ensemble average cluster size. Predicted trajectories from theory (Appendix \ref{sec:minimal_model:two_distant_clusters}) for three different time constants ($\tau=500, 1000, 2250~\mathrm{s}$) are shown by dashed lines. (B) Time-averaged strand size of dissolving simulations as a function of ripening time with error bars showing standard deviation of strand size over run. Black dotted line is a least-squares linear fit $\bars_{fit} = a  t_D + b$ with $a=.043$ and $b=58.6$. 
}
\label{figs:fig8}
\end{figure}

We find that ripening is the dominant mode of cluster evolution in these simulations ($p_r=63.9$).  The majority of dissolving events took place when the segment between two clusters consisted of only a few tens of beads, highlighting the total viscous drag acting on the connecting segment governs the rate at which one cluster absorbs chain from another.  During ripening, the segment between clusters moved superdiffusively in index space, while the cluster edges not adjacent to the segment (the outer segments) remained nearly fixed 
This behavior further supports the primary assumption of our minimal model for ripening—that the sizes of the outer pinned segments remain relatively unchanged compared to the segment between clusters.  

Ripening events occurred with seemingly random distribution. However, aligning all ripening events relative to the time of complete ripening, $t'=t - t_r$, reveals a consistent trajectory, as shown in Fig.~\ref{figs:fig8}. Our minimal model predicts that the cluster size near total ripening follows
\begin{equation}
\label{eq:length_time_dissolution}
    \ell(t) \approx \left(\ell_o^{4/3} - \frac{4 t}{3\tau}\right)^{3/4},
\end{equation}
to leading order, where $\tau = 3\,\pi \,\eta \,\ell_{tot}^{1/3}\, s /\gamma$ is the timescale set by the strand being incorporated into the larger condensate (see Appendix~\ref{sec:minimal_model:two_distant_clusters}).  Notably, the dissolving time, defined by $\ell(t_{diss}) = 0$, depends on the initial difference in cluster sizes. While the simulations were initialized with identical cluster sizes, our results indicate that fluctuations in cluster size are sufficient to create imbalances in the forces acting on clusters, ultimately driving ripening. This explains the distribution of dissolving events, as cluster size differences emerge through a stochastic process.


\section{Discussion}%
\label{sec:discussion}

In this study, we investigated how fundamental microscopic parameters governing
crosslinking influence the condensation of a constrained polymer. We
demonstrated a transition from a non-clustering to a clustering phase as a
function of binding affinity $K_e$ and polymer end-to-end separation $L_{sep}$.
Within the clustering phase, after initial nucleation and cluster growth, the
amount of chain incorporated into clusters reaches a steady state. Beyond this
point, the number of clusters decreases through two primary mechanisms: merging
(coalescence) and dissolving (ripening). Using a minimal model based on free
energy and mechanical arguments, we identified distinct timescales associated
with each mechanism. Our results demonstrate that the competition between these
two processes dictates the observed scaling behavior in our simulations.

Among the microscopic parameters examined, it is $L_{sep}$ and $K_e$ that primarily
determine macroscopic properties such as the maximum number of clusters and the
total chain length within clusters. In contrast, the turnover rate of
crosslinkers, $k_o$, does not affect these macroscopic properties or the
system’s scaling behavior but instead controls the rate at which these
quantities evolve. Our analysis suggests that varying the turnover rate
proportionally alters the timescales of merging and dissolving without affecting
their relative competition. Beyond a critical turnover rate, the system’s
dynamics become limited by the diffusive timescale associated with the motion of
clusters and chain segments within the viscous medium.  

To further investigate the interactions between clusters mediated by connecting
segments, we analyzed a system with two identically sized clusters separated by
a short connecting segment ($\sim6\%$ of the total chain length). Transient
fluctuations in cluster sizes created imbalances in the forces acting on the
clusters, which in turn triggered ripening. Unlike simulations that began
with unclustered chains, these systems predominantly exhibited ripening as the
primary mode of cluster evolution. Consistent with our minimal model and scaling
arguments, we found that the cluster ripening is linearly proportional
to the length of the segment connecting neighboring clusters. Moreover, we observe that
dissolving of a cluster follows a deterministic trajectory. Using our minimal model, we
provide a general ODE framework with a single free parameter that effectively
sets the timescale for this trajectory toward total dissolution. This timescale
is governed by the surface tension $\gamma$ of clusters in our phenomenological
model. While our study focuses on a specific sticky-tail polymer condensation
model, the underlying theory of cluster ripening applies broadly to any
constrained homopolymer with attractive interactions.  

These findings have broader implications for biological systems, particularly
chromatin organization within the nucleus. Given that cluster formation on a
constrained polymer chain can take tens of minutes to hours, such timescales may
be too slow for genomic processes such as transcriptional enhancement or
silencing. This raises the possibility that long-range hydrodynamic interactions
play a role in accelerating chromatin clustering
\cite{Zimm1956,Halperin2000,Klushin1998}. Indeed, active hydrodynamic forces
have been proposed to stir the nucleoplasm and enhance chromatin condensation
\cite{Saintillan2018,Zidovska2020,Mahajan2022}. A natural extension of our work
would be to incorporate these interactions and explore how different sources of
activity—such as forces generated by loop-extruding factors, polymerases, and
topoisomerases—affect chromatin organization. Furthermore, sequence-dependent
condensation and segregation have been observed in both \textit{in vitro} and
\textit{in vivo} studies \cite{Morin2022,Nguyen2022,Shrinivas2019}. Variations
in DNA-binding affinities could significantly influence cluster dynamics,
potentially accelerating collapse while ensuring that critical genes remain
transcriptionally active, thereby preventing unintended silencing.  

\section*{Acknowledgments}
The authors thank support from NSF grants DMR-2004469 and CMMI-1762506. The
computations in this work were performed at facilities supported by the
Scientific Computing Core at the Flatiron Institute, a division of the Simons
Foundation.

\begin{table}[h!]
  \caption{Summary of parameters for 3D simulations. System parameters are
  constant values used in simulations from previously known or measured values.
  Assay values were chosen to match experiments that employed $\lambda$-DNA
  \cite{Quail2021,Renger2022,Morin2022}. Microscopic model parameters determined
  the behavior of the sticky tails and existed within physiologically relevant
  ranges. These were scanned over to identify values at which condensates would
  form. Simulation parameters were chosen to provide long-time behavior without
  compromising accuracy. Parameters used in free-energy models are shown under
  coarse-grained parameters.}
  \label{tab:parameters}
  \centering
  \scriptsize
  \begin{tabular}{| l | l | S | l |}
  \hline
  \textbf{\normalsize Parameter} & \textbf{\normalsize Symbol} & \textbf{\normalsize Value} & \textbf{\normalsize Reference} \\
  \hline
  \textbf{System parameters} & & & \\
  \hline
  Bead (nucleosome) diameter & $b$ & \SI{10}{\nano\meter} & Well known \\
  \hline
  Diameter of amino acid & $b_{amino}$ & \SIrange{0.4}{1}{\nano\meter} & Well known \\
  \hline
  dsDNA persistence length & $l_p$ & \SI{50}{\nano\meter} & Well known \\
  \hline
  ssDNA persistence length & $l_{p,ss}$ & \SI{2.2}{\nano\meter} & Chi et al.\cite{Chi2013} \\
  \hline
  IDR persistence length & $l_{p,idp}$ & \SIrange{0.4}{0.5}{\nano\meter} & Hofmann et al.\cite{Hofmann2012} \\
  \hline
  Nucleotide length & $b_n$ & \SI{0.34}{\nano\meter} & Well known \\
  \hline
  Nucleotides in linker DNA & $N_{link}$ & 45 & Alberts et al.\cite{Alberts2017} \\
  \hline
  Double stranded DNA force plateau & $f_{ds}$ & \SI{65}{\pico\newton} &  Kumar et al. \cite{Kumar2010}\\
  \hline
  Maximum extension of dsDNA & $\epsilon_{max}$ & \num{1.7} & Kumar et al. \cite{Kumar2010} \\
  \hline
  DNA linker spring constant & $\kappa_{DNA}$ & \SI{4}{\pico\newton\per\nano\meter} & Calculated \\
  \hline
  Thermal energy & $k_B T$ & \SI{4.11}{\pico\newton\nano\meter} & Well known \\
  \hline
  Solvent viscosity & $\eta$ & \SI{1}{\pico\newton\second\per\micro\meter\squared} & Cytoplasm viscosity, Well known \\
  \hline
  \textbf{Assay values} & & & \\
  \hline
  Length of chain & $L$ & \SI{16.5}{\micro\meter} & Quail et al. \cite{Quail2021} \\
  \hline
  Number of beads & $N$ & 1650 & Calculated \\
  \hline
  Pinned distance & $L_d$ & \SIrange{2}{10}{\micro\meter} & Chosen \cite{Quail2021} \\
  \hline
  \textbf{Microscopic binding parameters}  & & & \\
  \hline
  Kinetic rate & $k_o$ & \SI{3}{\per\second} & Chosen (values range from 0.1–300) \\
  \hline
  Sticky end equilibrium constant & $K_e$ & \SI{30}{\per\micro\molar} & Chosen (values range from 0.1–300) \\
  \hline
  Binding site density & $\lambda_s$ & 1 & Subsumed into $K_e$ \\
  \hline
  Sticky end spring constant & $\kappa_{s}$ & \SI{800}{\pico\newton\per\micro\meter} & Calculated from FUS IDR values \cite{Kato2021} \\
  \hline
  Sticky end length & $l_X$ & \SI{0.005}{\micro\meter} & Calculated from FUS IDR values \cite{Kato2021} \\
  \hline
  \textbf{Simulation parameters} & & & \\
  \hline
  Timestep & $\Delta t$ & \SI{1e-4}{\second} & Chosen for accuracy, stability, and efficiency \\
  \hline
  \textbf{Coarse-grained parameters} & & & \\
  \hline
  Condensation free energy per volume & $\mu$ & \SIrange{4.1}{11.9}{\pico\newton\per\micro\meter} & Renger et al. \cite{Renger2022} \\
  \hline
  Condensate packing factor & $\alpha$ & \SIrange{0.05}{0.06}{\micro\meter\squared} & Renger et al. \cite{Renger2022}\\
  \hline
  Condensate surface tension & $\gamma$ & \SI{0.15}{\pico\newton\per\micro\meter} & Renger et al. \cite{Renger2022} \\
  \hline
  Filament flexibility & $k_BT/l_p$ & \SI{0.0822}{\pico\newton} & Calculated \\
  \hline
\end{tabular}
\end{table}

\bibliographystyle{plain}
\bibliography{dyn_cond_paper}

\newpage

\appendix
\section{Extended Simulation Methods, Protocols, and Parameters}%
All simulations were performed with with 1600 beads, each \SI{10}{nm} in
diameter and connected in series by spring-like bonds. The first and last beads
were held fixed to simulate being pinned to a coverslip, optically trapped, or
constrained by nuclear attachments
\cite{Zheng2018,Amiad-Pavlov2021,Pontvianne2016,
Wang1997,Kumar2010,Quail2021,Nguyen2022}. All simulations were performed using
the aLENS software package, which incorporates kinetic Monte Carlo binding of
bonds and excluded volume interactions between beads \cite{Lamson2021, Yan2022}.
The following subsections describe the initialization and setting of parameters
for each simulated assay.

\subsection{Sticky protein model}
\label{sub:sticky_protein_model}
The collapse of chains into clusters comes from the addition of transient crosslinks or ‘sticky tails’ to the beads composing the chains. In this section, we will discuss the interactions of beads through sticky tails and the rationale/theory behind the model.

To model attractive interactions between IDRs and other proteins, a sticky tail can bind to nearby beads, forming a crosslink between the original bead it was attached to and a neighboring bead. The aLENS software models crosslinking of proteins to other proteins through a kinetic Monte Carlo scheme that satisfies the detailed-balance condition. For a two-state system, this means
\begin{equation}
\label{eq:detailed-balance}
    \frac{k_{2\to 1}}{k_{1\to 2}} = \frac{P_1}{P_2} = \exp\left(\frac{G_{2} -G_{1}}{k_B T}\right),
\end{equation}
where $P_i$ is the probability of being in state $i$, $k_{i\to j}$ is the rate of state $i$ going to state $j$, and $G_i$ is the free energy associated with being in state $i$. Sticky tails are objects that take one of two main states: non-crosslinking (0) and crosslinking (1). In the non-crosslinking state, the sticky tail has  free energy $G_0 = \epsilon_0 - k_BT \ln{Z_o}$ while in the unbound state. The non-crosslinking partition function accounts for all configurations of a spring while attached to a bead: 
\begin{equation}
\label{eq:Z_0}
    Z_0 = \frac{1}{V}\int_V dr \exp\left(-\frac{U(\br)}{k_B T}\right)=\frac{V_0}{V},
\end{equation}
where $U(r)$ is the energy stored in the tail when stretched a distance $r$ from the center of the bead and $V$ is the volume of the domain. Physically, $V_0$ is the weighted volume that the sticky tail explores. For stiff and short springs, $V_0$ is small, concentrating the volume in which the tail will bind. We assume the tail explores a spherically symmetric space around the bead's center; without loss of generality, we may set the energetic offset $\epsilon_0=0$. 

In the crosslinking state, the sticky tail no longer has an entropic contribution, but we must take into account the tail's energy while stretched as well as the energy involved in being bound to another object. The crosslinking energy is then $G_1(|\br_i - \br_j|) = \epsilon_1 + U(|\br_i - \br_j|)$, where $\br_i$ is the position of bead $i$. The potential spring energy now accounts for the fact that the tail is being stretched over the distance separating the two crosslinked beads $i$ and $j$. Since $G_0$ is constant, we can use equations \eqref{eq:detailed-balance} and \eqref{eq:Z_0} to calculate the probability that a sticky tail is not crosslinked:
\begin{equation}
P_0(\br_j) = \left(1+\frac{K_e}{V_0}\sum_i^n\lambda \exp{\left(-\frac{U(|\br_i - \br_j|)}{k_BT}\right)}\right)^{-1},
\end{equation}
where $K_e = V \exp{\left(-\epsilon_1/k_BT\right)}$ is known as the equilibrium constant for the reaction between the sticky tail and a bead and $\lambda$ is the binding site multiplicity on a bead. In our simulations, we will combine the values of $K_e$ and $\lambda$.

To use the kinetic Monte Carlo scheme, we need to express these values as a rate, which is done by assuming that the unbinding rate of the tail $k_{ij \to i} = k_o$ is constant. Using equation \eqref{eq:detailed-balance}, we can write down the binding rate of a tail fixed to bead $i$ attaching to bead $j$ as 
\begin{equation}
k_{i\to ij} = \frac{k_o K_e}{V_0} \exp{\left(-\frac{U(|\br_i - \br_j|)}{k_BT}\right)}.
\end{equation}
From this rate  we are able to stochastically sample those tails that unbind and
bind to beads in a way that satisfies principles of statistical mechanics. In
our model, we consider the sticky tails to be responding like Hookean springs
and therefore use the potential $U(\br) =
\frac{\kappa}{2}\left(|\br|-r_0-b\right)^2$, where $\kappa$ is the spring
constant and $r_0$ is the rest length of the tail. Since we bind to the center
points of beads, we include the bead diameter $b$ in our calculations to account
for the fact that the tail is fixed and binding to surfaces of the bead and not
the center.

We perform a version of kinetic Monte Carlo to sample the proper binding kinetics at every instance. Our algorithm considers every tail individually, calculating the probability of undergoing a state change. For non-crosslinking tails, we calculate  crosslinking rates to the set $N_b$ of neighboring beads. We impose a distance cutoff based on whether the equilibrium probability of two beads being bound is greater than $P_e = 10^{-5}$ i.e. $r_{cut} = l_X + \sqrt{2k_B T\ln(K_e/P_e)/\kappa_X}$.
We then invoke the fundamental premise of stochastic chemical kinetics to calculate the probability of binding to these neighboring beads. The temporal probability density that a tail initially attached to bead $i$ will crosslink to bead $j$ at time $t$ is given by
\begin{equation}
\label{eq:pij}
    p(ij,t;i,0) = k_{i\to ij} \exp{\left(-t\sum_l^{N_b} k_{i\to il}\right)},
\end{equation}
The probability that a reaction occurs in a timestep $\Delta t$ is then
\begin{equation}
\label{eq:Pij}
    P(ij,\Delta t;i,0) = \int_0^{\Delta t} dt\, p(ij,t;i,0)=\frac{k_{i\to ij}}{\sum_l^{N_b} k_{i\to il}}\left( 1- \exp{\left(-\Delta t\sum_l^{N_b} k_{i\to il}\right)}\right).
\end{equation}
One can then calculate the probability of no event occurring as
\begin{equation}
    P(i,\Delta t;i,0) = 1-\sum_l^{N_b}P(il,\Delta t;i,0)= \exp{\left(-\Delta t\sum_l^{N_b} k_{i\to il}\right)}.
\end{equation}
For tails that have already crosslinked, the only transition rate is the constant unbinding rate. Using the principles behind equations \eqref{eq:pij} and \eqref{eq:Pij} for a single rate, we can write down the probability of unbinding from bead $j$ as 
\begin{equation}
    P(i,\Delta t;ij,0) = 1 - \exp{(-k_o \Delta t)}.
\end{equation}

The sticky tail binding algorithm is then 
\begin{algorithm}
\caption{Sticky tail binding algorithm}
\begin{algorithmic}[1]
\FOR{every tail attached to bead $i$}
    \STATE Sample uniform random number $r \gets U[0,1]$ 
    \IF{tail is crosslinked}
        \STATE Calculate unbinding probability $P(i,\Delta t;ij,0) \gets 1 - \exp{(-k_o \Delta t)}$
        \IF{$r < P(i,\Delta t;ij,0)$}
            \STATE Unbind tail
        \ENDIF
    \ELSE
      \STATE Calculate set of neighboring beads $N_b$
      \STATE Calculate binding rates $k_{i\to ij}\gets\frac{k_o K_e}{V_0} \exp{\left(-\frac{U(|\br_i - \br_j|)}{k_BT}\right)}$
      \STATE Calculate sum of binding rates $z_{tot} \gets \sum_l^{N_b} k_{i\to il}$
      \STATE $P_{sum} \gets 0$
      \FOR{every bead $j$ neighboring $i$}
          \STATE Calculate binding probability $P(ij,t + \Delta t;i,t) \gets k_{i\to ij} \frac{1-\exp{(-z_{tot}\Delta t )}}{z_{tot}}$ 
          \STATE $P_{sum} \gets P_{sum} + P(ij,t + \Delta t;i,t)$
          \IF {$r < P_{sum}$}
              \STATE Bind tail to bead $j$
              \STATE \textbf{break}
          \ENDIF
      \ENDFOR
    \ENDIF
\ENDFOR
\end{algorithmic}
\end{algorithm}

Once crosslinked, sticky tails exert a force between the two crosslinked beads consistent with the potential energy used to calculate the binding probability:
\begin{equation}
    \bF_{ij} = -\kappa\left(1-\frac{r_0+b}{|\br_i - \br_j|}\right)(\br_i - \br_j),
\end{equation}
where $\bF_{ij}$ is the force exerted on bead $i$ by bead $j$.

\subsection{Parameter Calculations}
\label{sub:parameter_calculations}

\paragraph*{Bead diameter}
The bead diameter was chosen to match the size of a nucleosome. The length of the linker DNA was only taken into account in the calculation of the spring constant $\kappa_{DNA}$. The number of nucleotides per histone is 147, which corresponds to an unstretched DNA length of $\approx \SI{50}{\nano\meter}$ since a nucleotide is \SI{.34}{\nano\meter} long. 

\paragraph*{Linker spring constant}
DNA force-extension curves, measured from experiments, have a non-trivial shape.  Between strains 0 and 1, the polymer behaves like a worm-like chain model. The force extension relation for this model is
\begin{equation}
 f(\epsilon) = \frac{k_{B}T}{l_p} \left( \frac{1}{4(1-\epsilon)^2} -\frac{1}{4} + \epsilon \right),   
\end{equation}
where $\epsilon$ is the separation of DNA ends divided by DNA contour length, $k_{B}T$ is the thermal energy, and $l_p$ is the persistence length (Table \ref{tab:parameters}). Experiments have shown that as the strain approaches 1, the DNA is stretched to the point where the double helix begins to unwind and the force plateaus at a value of 65 pN \cite{Kumar2010}.  Beyond a strain of 1.7, the DNA  may break, leaving only one strand of DNA exerting force. Single-stranded DNA also follows a WLC force extension curve, although with a smaller persistence length. 
\begin{figure}
    \centering
    \includegraphics[width=0.7\linewidth]{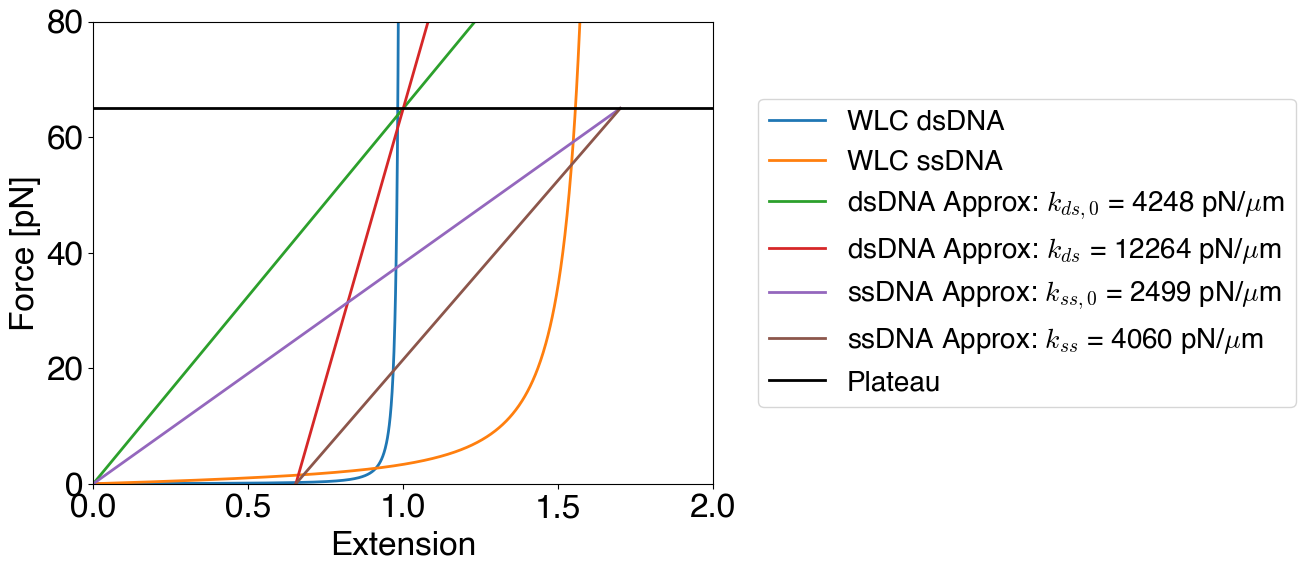}
    \caption{Comparison of worm-like chain and linear spring models for links between chain monomers. Spring constants for linear springs are chosen to match the unwinding force of DNA (65 pN) at extensions of 1 and 1.7. Rest length is chosen to be either zero (linker DNA bound to surface of nucleosomes) or 1 bead diameter (linker DNA bound to center of nucleosomes).}
    \label{sfig:DNA_force_extension}
\end{figure}

In our code, we consider linear springs with a non-zero rest length as shown in  Fig.~\ref{sfig:DNA_force_extension}. To find equivalent parameters, we assume that beads (nucleosomes) prefer to be separated by one bead diameter. From the WLC model, we obtain the maximum distance linker DNA could span from bead centers. It is then possible to solve for the spring constant $\kappa_{DNA}$ in the equation $f_{ds} = \kappa_{DNA}(N_{link} l_n \epsilon_{max} - b)$, where $f_{ds}$ is the force at which the DNA begins to unwind, $N_{link}$ is the number of nucleotides in the linker DNA, $l_n$ is the length of a nucleotide, and $\epsilon_{max}$ is the maximum extension of the DNA before it breaks ($\epsilon_{max}=1.7$).  We find $\kappa_{DNA} = 4060$ pN/$\mu \text{m}$ (brown line). If we assume that the linker DNA never stretches beyond its contour length, then $f_{ds} = \kappa_{DNA}(N_{link} l_n - b)$, which increases the spring constant to $\kappa_{DNA} = 12264$ pN/$\mu \text{m}$ (red line). If one considers the extension to be measured from the surfaces of beads, the green and purple curves are obtained. 

Our goal here is to determine approximate values for the properties of chromatin. It is therefore reasonable to assume that the spring constant of the linker DNA is somewhere between 2500 pN/$\mu \text{m}$ and 12500 pN/$\mu \text{m}$. We choose a value of 4000 pN/$\mu \text{m}$ (Table \ref{tab:parameters}).


\subsection{Initialization of systems}
 To mitigate biases introduced from initial conditions, all simulations were started from the last configuration of equilibrated constrained chains. Two different types of assays were performed, which required different initializing methods: nucleating cluster assays and preformed cluster assays.
 
\subsubsection{Initializing nucleating cluster simulations}
\label{ssub:nucleating_cluster_simulations}
Nucleating cluster simulations (Figs. \ref{figs:fig1} - \ref{figs:fig6})
replicate experiments and scenarios where DNA or other flexible biofilaments
have just been populated by crosslinking and/or attractive proteins. As such,
the filaments should start out in probable equilibrium configurations of
self-avoiding polymer chains and then evolve under the presence of proteins. 

To equilibrate chains before introducing crosslinking elements, beads were
placed touching each other along a helical curve with $p = 10$ rotational
periods. Each helix spans a distance $L_{sep}$ with an arc length of
$L_{tot}=16\mu\text{m}$ while maintaining constant curvature. This removed
artificially prescribed regions with higher chances of clustering. Hookian
spring bonds connected neighboring beads, creating an unbroken chain.

After initial placement, chain displacements were evolved for 30 seconds while
the end beads were held fixed in place. Only steric, chain bond, and Brownian
forces influenced the bead motion during this step. Twelve replicates were
created, each with a different random number generator seed, producing different
equilibrated end configurations. 

The final bead and bond configurations from the above simulations were used as
the initial configurations for the nucleating cluster assays. In addition, every
bead was populated with two ‘sticky tails’ (see section
\ref{sub:sticky_protein_model}. Equations of motion were then integrated using
aLENS's hybrid time-stepping algorithm that incorporates sticky tail binding
kinetics, chain and tail bond forces, and steric interactions while still fixing
the end beads in space.

\subsubsection{Initializing preformed cluster simulations}
\label{ssub:preformed_cluster}

Preformed cluster simulations (Figs. \ref{figs:fig7} - \ref{figs:fig8}) were
initialized to study the dynamics of two clusters of equal size and the effect
of the distance between had on the clusters' growth dynamics. The initial configuration of the
system was created by placing two clusters of equal size a distance of $s$
apart. 

As described in the previous section, the initialization process began with
beads placed in a helical configuration where end beads were held fixed at
$L_{sep} = 5 \mu m$. We then equilibrated the bare chain for 30 seconds. From
here, two clusters of equal size were formed by placing sticky ends with $K_e=30
\mu M^{-1}$ on a subset of beads that consisted of two contiguous regions. The
total number of beads with sticky ends was chosen to equal the steady-state
total clustered lengths seen in Fig.~\ref{figs:fig2} ($\ell_{tot} = 9.6
\mu\text{m}$). For the preformed cluster runs, regions were separated by 100
beads (1 $\mu\text{m}$). After running the simulation for 60 seconds, the system
would have two fully condensed clusters (Fig.~\ref{figs:fig7}). Lastly, the
remaining backbone of the chain was populated with crosslinkers creating the
final initial condition so that clusters could translocate, grow, merge, and
dissolve.

\section{Minimal model of cluster dynamics}%
\label{sec:minimal_model}

\begin{figure}[h]
  \centerline{\includegraphics[width=0.8\textwidth]{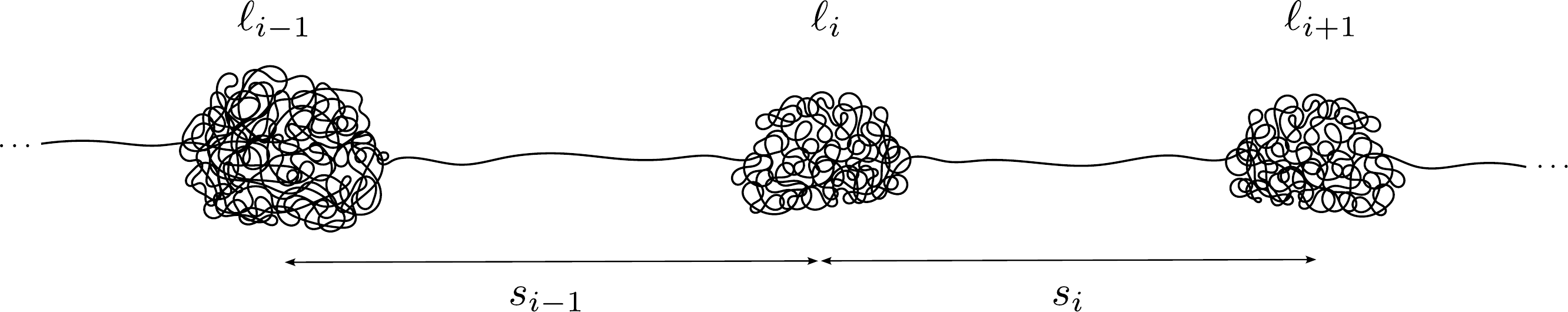}}
  \caption{Schematic representation of the $i$-th cluster with length $\ell_{i}$, separated from its neighboring clusters by two stretched segments of lengths $s_{i-1}$ and $s_{i}$ on its left and right, respectively.}\label{figs:two_clusters}
\end{figure}

We begin by considering the total length of chain $L_{tot}$, partitioned into
clusters with length $\ell_i$ and connecting strands of length $s_i$
(Fig.~\ref{figs:two_clusters}). Assuming each cluster has two connecting strands
and that no loop strands are present, the total length of the polymer can be
expressed as:
\begin{equation}
L_{tot} = \ell_{tot} + s_{tot}, \label{eq:elltot_stot_Ltot}
\end{equation}
where $\ell_{tot} = \sum_i^{N_c} \ell_i$ and $s_{tot} = \sum_i^{N_c+1} s_i$ and
$N_c(t)$ is the number of clusters at a given time $t$. We introduce the
notation $\langle \dots \rangle$ to denote an ensemble average of any given
quantity. In section \ref{sub:determinant_of_clustering}, we observed that for
all cluster-forming parameter sets, $\AVG{\ell_{tot}}$ has a standard deviation
of less than $3$\% after the initial nucleation stage, $t\ge t_{sat}$, where
$t_{sat}$ marks the onset of the plateau in $\ell_{tot}$ as seen in
Fig.~\ref{figs:fig2}(A). Thus, we consider $\ell_{tot}$ to be approximately
constant in this regime. This assumption implies that $s_{tot} = L_{tot} -
\ell_{tot}\approx Const$ for $t\ge t_{sat}$. The values of $s_{tot}$ and
$\ell_{tot}$ are determined by the parameters $K_e$ and $L_{sep}$.

We now consider the average length of chain in clusters and strands: $\barell=\ell_{tot}/N_c$ and $\bars=\frac{s_{tot}}{N_c+1}\approx s_{tot}/N_c$, respectively. From these equations, we see that the average cluster size and strand size are related by $\bars = a \,\barell$, where $a=(s_{tot}/\ell_{tot})$. Fig.~\ref{figs:fig2}(C) shows that the total length of clustered chains $\ell_{tot}$ in the steady-state decreases almost linearly with $L_{sep}$:
\begin{equation}
\ell_{tot} \approx L_{tot} - c\, L_{sep},\qquad  \text{for}~L_{sep}< L_{tot}, \label{eq:elltot_Lsep}
\end{equation}
where $c$ is the slope measured from Fig.~\ref{figs:fig2}(C). Comparing equations \eqref{eq:elltot_Lsep} and \eqref{eq:elltot_stot_Ltot}, we note that $s_{tot} \approx c\,L_{sep}$, and therefore
\begin{equation}
    a = \frac{s_{tot}}{\ell_{tot}} \approx \left({\frac{L_{tot}}{c\,L_{sep}}-1}\right)^{-1},
\end{equation}
which is a constant set by the simulation parameters $L_{tot}, K_e$ and $L_{sep}$ as shown in Fig.~\ref{figs:fig2}.

After nucleating a number of clusters, the clusters may disappear or grow through ripening (dissolving) or coarsening (merging) mechanisms. The corresponding kinetic equation describing the evolution of the number of clusters is:
\begin{equation}
  \label{eq:dot_Nc}
  \dot{N}_c = -(k_{diss} + k_{merge}) (N_c-1),\qquad N_c\ge1, 
\end{equation}
where the merging rate $k_{merge}= \tau_{merge}^{-1}$ and dissolving rate
$k_{diss}= \tau_{diss}^{-1}$ are both functions of $N_c$ and are determined by
the current configuration of the system. The reason for  multiplying the rates
on the right-hand side of equation~\eqref{eq:dot_Nc} by $N_c-1$ instead of $N_c$
is to account for the two strands attached to the end points, which do not
contribute to the dissolving or merging events. Note that this enforces the
limit, $ \lim_{t\to\infty} (N_c,\,\dot{N}_c)=(1,\,0)$.  In sections
\ref{sec:minimal_model:dissolve_rate} and \ref{sec:minimal_model:merging_rate},
we determine these rates as a function of the macroscopic variables in the
system, including $N_c$, $\ell_{tot}$, and $\langle \delta \ell_i^2 \rangle$.
Next, in section \ref{sec:minimal_model:cluster_number_avg_length}, we discuss
the scaling behavior in different regimes where merging or dissolving is
dominant. In addition, we are able to relate the change in cluster size to the
number of of clusters:
\begin{align}
&\bar{\ell}= \frac{\ell_{tot}}{N_c },\\
&\frac{\dot{\barell}}{\barell^2} = -\frac{\dot{N}_c}{\ell_{tot}}.\label{eq:l_to_N}
\end{align}
The variance of the cluster sizes $\langle \delta \ell_i^2 \rangle$ is empirically determined from simulations, which shows a near constant value as the system evolves.

\subsection{Ripening (Dissolving) Rates} \label{sec:minimal_model:dissolve_rate}
For a cluster to dissolve, a bead must detach and leave that cluster. If
$\ell_{tot}$ remains constant, a different bead must subsequently join another
cluster, effectively transferring beads from the $i$-th cluster to the $j$-th
cluster.  We note that this could be a non-local transfer in which cluster $i$
may decrease in size while a non-neighboring cluster $j$ increases. However,
this would require all clusters and strands in between the two changing clusters
to spatially translocate to accommodate the rearrangement. Balancing forces, we
write down the velocity of all strands and clusters between clusters $i$ and $j$
($i<j$) as:

\begin{align}
  &v_{ij} =  \mu_{ij} \, f_{ij} ,\label{eq:velocity_force_mobility}\\
  &\mu_{ij} \cong \left(\eta \sum_{n=i}^{j-1} (l_{n+1}+s_n)\right)^{-1}, \label{eq:mobility_muij}
\end{align}
where $f_{ij}$ is the net force exerted on strand $i$ by cluster $i$ and on
strand $j-1$ by cluster $j$,  $\mu_{ij}$ is the mobility that accounts for the
viscous drag on all beads between clusters $i$ and $j$\footnote{$\cong$ implies
that the equality holds up to a purely numerical prefactor,  whereas $\sim$
implies a proportional relationship where the prefactor could contain a
dimensional value. $\approx$ denotes the absence of higher-order terms of the
functional relationship.}, and $\eta$ is the solvent viscosity. We demonstrate
in the following that this velocity determines the rate of cluster ripening
\cite{Klushin1998}. Based on equation \eqref{eq:mobility_muij}, we know that
these strands possess the largest mobility coefficients, as $\mu_{i, i\pm 1} >
\mu_{i, i\pm n}$ for $n>1$ . Therefore, for a given $f_{ij}$, the largest
velocities will belong to the strands connecting neighboring clusters.
Consequently, we can neglect forces exerted by non-neighboring clusters and
approximate the average rate of ripening as:

\begin{equation}
  \label{eq:k_diss}
  k_{diss} \cong \frac{\langle |v_i|\rangle}{\barell}\cong\frac{\langle|f_{i,i+1}|\rangle}{\eta \bars \barell}.
\end{equation}
The average magnitude of the net force can be approximated as
\begin{equation}
    \langle|f_{i,i+1}|\rangle \geq \sqrt{\langle f_{i,i+1}^2\rangle},
\end{equation}
which is equivalent to the variance of the forces along the polymer. The
directional force generated by cluster $i$ on its adjacent strand $i$ is
described by 
\begin{equation}
    \label{seq:cluster_force}
    f_i = -\alpha + \gamma \ell_i^{-1/3},
\end{equation}
 where $\alpha$ represents the cluster's bulk energy and $\gamma$ corresponds to the surface tension. This form reflects the interplay between bulk and surface energy contributions as a function of cluster size, and is derived from scaling analyses that connect the force to the size of the cluster  \cite{Halperin1991}.

Next, we calculate the net force between neighboring clusters as a function of deviations from the mean cluster size $\barell$:
\begin{align}
\label{eq:Delta_f}
f_{i,i+1} &= \gamma \left((\barell + \delta\ell_i)^{-1/3} - (\barell + \delta\ell_{i+1})^{-1/3}\right)\\
                &= \gamma \barell^{-1/3}\left((1 + \varepsilon_i)^{-1/3} - (1 + \varepsilon_{i+1})^{-1/3}\right),
\end{align}
where $\varepsilon_i = \delta \ell_{i}/\barell$, with both $\varepsilon_i$ and $\delta \ell_i$ treated as random variables satisfying $\langle \delta \ell_i\rangle = 0$, $\langle \delta \ell_i \delta \ell_j \rangle = \langle \delta \ell_i \rangle \langle \delta \ell_j \rangle = 0$. As discussed in Section~\ref{sec:minimal_model}, the variance of cluster sizes, normalized by $\barell^2$, remains approximately constant once the simulations reach the maximum $\ell_{tot}$. Thus, we can assume $\langle \varepsilon_i^2\rangle=\langle\delta\ell_i^2\rangle/\barell^2\approx Const$ over time (see Fig.~\ref{figs:fig3}(C,F)). To approximate the variance of the force, we expand $f_{i,i+1}$ under small deviations $\varepsilon \ll 1$ and use the second moment to derive a relationship for $\dot{\barell}$:
\begin{align}
    f_{i,i+1}^2 &= \gamma^2 \barell^{-2/3} \left(-\frac{1}{3}\varepsilon_i + \frac{1}{3}\varepsilon_{i+1}\right)^2 +O(\varepsilon^3)\\
    &\approx \gamma^2 \barell^{-2/3}\left(\frac{1}{9}\varepsilon_i^2 + \frac{1}{9}\varepsilon_{i+1}^2-\frac{2}{9}\varepsilon_{i}\varepsilon_{i+1}\right).
\end{align}
The ensemble average of the force magnitude is given by
\begin{equation}
    \langle|f_{i,i+1}|\rangle \approx \sqrt{\langle f_{i,i+1}^2\rangle} =\frac{1}{3} \gamma \barell^{-1/3}
    \sqrt{2\langle\varepsilon_i^2\rangle + O(\langle \varepsilon_i^3\rangle)}. \label{eq:force_mag_average}
\end{equation}
Finally, we obtain the dissolving rate by plugging equation \eqref{eq:force_mag_average} into \eqref{eq:k_diss}:
\begin{equation}
    \label{eq:k_diss_2}
    k_{diss} \cong\frac{\gamma}{\eta \bars \barell^{4/3}} \sim \barell^{-7/3}\quad \text{or} \quad \tau_{diss} \cong \barell^{7/3}.
\end{equation}

\subsection{Coalescence (Merging) rate}
\label{sec:minimal_model:merging_rate}
In this section, we determine the  merging rate of two neighboring clusters, occurring on a timescale associated with one-dimensional diffusion along the connecting strand:
\begin{equation}
    \tau_{merge} \cong s^2/D_c. \label{eq:tau_merge}
\end{equation}
Here,  $D_c \cong k_BT/\eta \ell$ is the diffusion constant of a cluster. Therefore, the merging rate is given by 
\begin{equation}
    \label{eq:k_merge}
    k_{merge} \cong \frac{k_BT}{\eta \bars^2 \barell} \sim \barell^{-3}.
\end{equation}
The merging rate predicted in equation \eqref{eq:k_merge} is significant as it establishes a bounding behavior for our simulations. This behavior is relevant when $t-t_{sat}$ is relatively small and the clusters are relatively close. However, at longer times, as clusters drift farther apart, ripening becomes the dominant dynamic.

\subsection{Number of clusters and average cluster size dynamics}\label{sec:minimal_model:cluster_number_avg_length}
We can now solve for the average number of clusters and the average cluster size as a function of time. We start by substituting equations \eqref{eq:k_diss} and \eqref{eq:k_merge} into \eqref{eq:dot_Nc}:
\begin{equation}
  \label{eq:dot_Nc_sub1}
  \dot{N}_c =  -\left(\tilde{c}_1 \frac{\gamma}{\eta \bars \barell^{4/3}} + \tilde{c}_2\frac{k_B T}{\eta \bars^2 \barell} \right)(N_c-1),
\end{equation}
where $\tilde{c}_1$ and $\tilde{c}_2$ are constant coefficients for $k_{diss}$ and $k_{merge}$. Gathering constants and expressing all terms as functions of $N_c$, we obtain:
\begin{align}
  &\dot{N}_c =  - c_1 \, \big(\underbrace{\frac{a k_BT}{\eta L_{sep}^3}}_{\bar{\tau}^{-1}} \big) \Bigg[ c_2 \, \big(\underbrace{ \frac{\gamma a^{1/3} L_{sep}^{2/3}}{k_B T}}_{A}\big) \,N_c^{7/3}  + N_c^3 \Bigg](N_c-1), \label{eq:dot_Nc_sub2}\\
  & c_1 =\tilde{c}_2/c^3, \qquad c_2={\tilde{c}_1\, c^{2/3}}/{\tilde{c}_2}, \label{eq:dot_Nc_sub2_c1_c2}
\end{align}
where we have used the relation  $\ell_{tot} = c\, L_{sep}/a$. In our
simulations, we observe approximate values of  $c \approx 1$ ,  $\varepsilon
\approx 1$, $c_1 \approx \sqrt{2} \varepsilon / 9\pi$, and $c_2 \approx
(3\pi)^{-1}$. From equation \eqref{eq:dot_Nc_sub2}, we obtain a timescale
$\bar{\tau} = {\eta L_{sep}^3}/{k_BT a}$ and a dimensionless parameter $A
={\gamma a^{1/3} L_{sep}^{2/3}}/{k_B T}$. Notably, the timescale increases with
the end-to-end separation $L_{sep}$, consistent with our with observation in
Fig.~\ref{figs:fig2}(A). The ratio $A\, N_c^{-2/3}$ indicates the relative rates
of dissolving and merging, determining the dominant mode of cluster evolution
during filament dynamics. When $A\, N_c^{-2/3}<1$, merging dominates;
however, when $A\, N_c^{-2/3}>1$ as the number of cluster decreases, dissolving
becomes more frequent. Explicitly solving for $N_c$ in regimes where only one
mechanism dominates and $N_c \gg 1$,  equation \eqref{eq:dot_Nc_sub2} simplifies
to
\begin{align}
&\text{ripening}:& &\dot{N_c} = -\frac{c_1 c_2 A}{\bar{\tau}}N_c^{10/3} & 
\Rightarrow N_c \sim t^{-3/7} \\
&\text{merging}:& &\dot{N_c} = -\frac{c_1}{\bar{\tau}}N_c^{4} &
\Rightarrow N_c \sim t^{-1/3}.
\end{align}
These are the bounding power laws seen in Fig.~\ref{figs:fig3}. 

Using equation \eqref{eq:l_to_N}, we reformulate \eqref{eq:dot_Nc_sub2} in terms of the average cluster length $\barell$ as
\begin{equation}
  \label{eq:final_dot_ell}
  \frac{\dot{\barell}}{\ell_{tot}} = \frac{c_1}{\bar{\tau}} \, \left[ c_2\, A
  \,\left(\frac{\barell}{\ell_{tot}}\right)^{-4/3} +
  \left(\frac{\barell}{\ell_{tot}}\right)^{-2} \right]
  \left(1-\frac{\barell}{\ell_{tot}}\right). 
\end{equation}
Equation \eqref{eq:final_dot_ell} predicts that for $\barell < \ell_{tot}$,
cluster sizes continuously increase but the rate of increase slows down over
time. This is consistent with our simulations, as shown in
Fig.~\ref{figs:fig2}(A).

\subsection{Maximum number of clusters}
\label{sec:minimal_model:maximum_cluster_number}
To determine the maximum number of clusters that can coexist in an unstable
equilibrium, we assume a chain configuration in which equally sized clusters are
uniformly spaced along the chain. Due to Newton's 3rd law, the force exerted by
clusters on the neighboring strands counterbalances the force exerted by the
strands on the clusters: $0 = f_i + f_{p}$, where $f_i$ is the force exerted by
a single condensate in one strand and $f_p$ is the force exerted by a strand on
its neighboring cluster. At equilibrium, clusters remain stationary, with the
worm-like chain exerting a restorative force under stretching, given by
\begin{equation}
    f_{wc}(x) = \frac{\kappa x^2}{4}\left(\frac{1}{\left(s - x\right)^2} + \frac{2}{s^2}\right),\label{eq:force_wormlike_chain}
\end{equation}
where $\kappa$ is the chain's flexibility, $s$ is the length of the polymer and
$x$ is the end-to-end extension. For multiple clusters regularly spaced along
the backbone of a chain, the average distance between clusters is $\bars =
\frac{L_{tot}}{N_c} - \barell$ and the end-to-end extension for a single strand
is approximately $\bar{x}=L_{sep}/N_c - \barell^{1/3} \approx L_{sep}/N_c$. To
determine the maximum number of clusters that can satisfy the force balance, we
start by combining equations \eqref{seq:cluster_force} and
\eqref{eq:force_wormlike_chain}:
\begin{equation}
    0 = -\alpha + \gamma \barell^{-1/3} + \frac{\kappa L_{sep}^2}{4} \left(\frac{1}{\left(L_{tot} - L_{sep} - N_c\barell\right)^2} + \frac{2}{(L_{tot}-N_c\barell)^2}\right).
\end{equation}
We neglect the second term in the expression for polymeric force, as it is minimal when the chain is significantly stretched. Solving for $N_c$, we find:
\begin{equation}
    N_c(\barell) =\frac{1}{\barell}\left[L_{tot} - L_{sep}\bigg[ 1 + \frac{1}{2}\bigg(\frac{\kappa}{\alpha - \gamma \barell^{-1/3}} \bigg)^{1/2} \bigg] \right],
\end{equation}
indicating that the maximum number of clusters in equilibrium depends on the
cluster size $\bar{\ell}$. Maximizing this function results in a monotonically
decreasing dependence on $L_{sep}$, consistent with the trend observed in
Fig.~\ref{figs:fig3}(D).

\subsection{Two distant clusters}
\label{sec:minimal_model:two_distant_clusters}
In this section, we focus on two clusters of size $\ell_1$ and $\ell_2$ that are
separated by a strand of length $s$. We consider the case where $s\gg k_B T
{\bar{\ell}}^{1/3}/\gamma$. In this limit, ripening processes dominate over
merging, since $k_{diss}\gg k_{merge}$. This scenario aligns with the behavior
illustrated in Fig.~\ref{figs:fig8}. At long times, when the total length of the
chain in clusters reaches a steady state, the cluster lengths satisfy
\begin{equation}
\ell_2 = \ell_{tot} - \ell_1. \label{eq:2_cluster_length_conservation}
\end{equation}
Using equations \eqref{eq:velocity_force_mobility} and \eqref{seq:cluster_force},
the rate of dissolving for the first cluster can be estimated as \begin{align}
    \dot{\ell}_{1} &= -v_s= -\mu_{12} \,f_{12} \\
                &= -\frac{\gamma}{3 \pi \eta s} \left( \ell_1^{-1/3} - \ell_2^{-1/3}\right).
\end{align}
We define $\tell = \ell_1/\ell_{tot}$ and $\ell_2/\ell_{tot} = 1- \tell$ by
non-dimensionalizing the cluster sizes by $\ell_{tot}$ and using equation
\eqref{eq:2_cluster_length_conservation}. Consequently, the rate of change of
the normalized size for a single condensate is given by
\begin{equation}
    \dot{\tell} = -\frac{1}{\tau} \left( \tell^{-1/3} - (1-\tell)^{-1/3}\right), \label{eq:two_cluster_dot_ell}
\end{equation}
where $\tau = 3\pi \eta s\,\ell_{tot}^{4/3} /\gamma$ is the characteristic
timescale of dissolving. An analytic solution for $t$ as a function of $\tell$
in  equation \eqref{eq:two_cluster_dot_ell} exists in terms of hypergeometric
functions. This allows for a solution for the time it takes for clusters to
dissolve with a given a cluster asymmetry. The curves in Fig.~\ref{figs:fig8}
are bounded by the numerically integrated solutions of equation
\eqref{eq:two_cluster_dot_ell} with $\tau \in [500, 2250]$ seconds.

\section{Cluster Definition and Tracking Algorithm}%
\label{sub:appendix:cluster_def_tracking}
To analyze  cluster dynamics in our simulations, we must first define what a
cluster is. Previous techniques have included the use of heuristics such as
density, connectivity, or contact map analysis \cite{Majumder2015, Sevick1988,
Lappala2013a}. Here we have chosen to use an unsupervised machine learning
method known as density-based spatial clustering of applications with noise
(DBSCAN) \cite{Ester1996} to identify and define clusters. After classifying
which beads are in clusters for each time point, we then temporally associate
clusters based on a similarity score and reconstruct the history using a
modified version of the \verb|SUBLINK| \cite{Rodriguez-Gomez2015}, a recursive
algorithm used to construct subhalo merger trees of simulated galaxies. 

\subsection{DBSCAN cluster identification}
\label{sub:appendix:cluster_def_tracking:DBSCAN}
\begin{figure}
    \centering
    \includegraphics[width=0.8\linewidth]{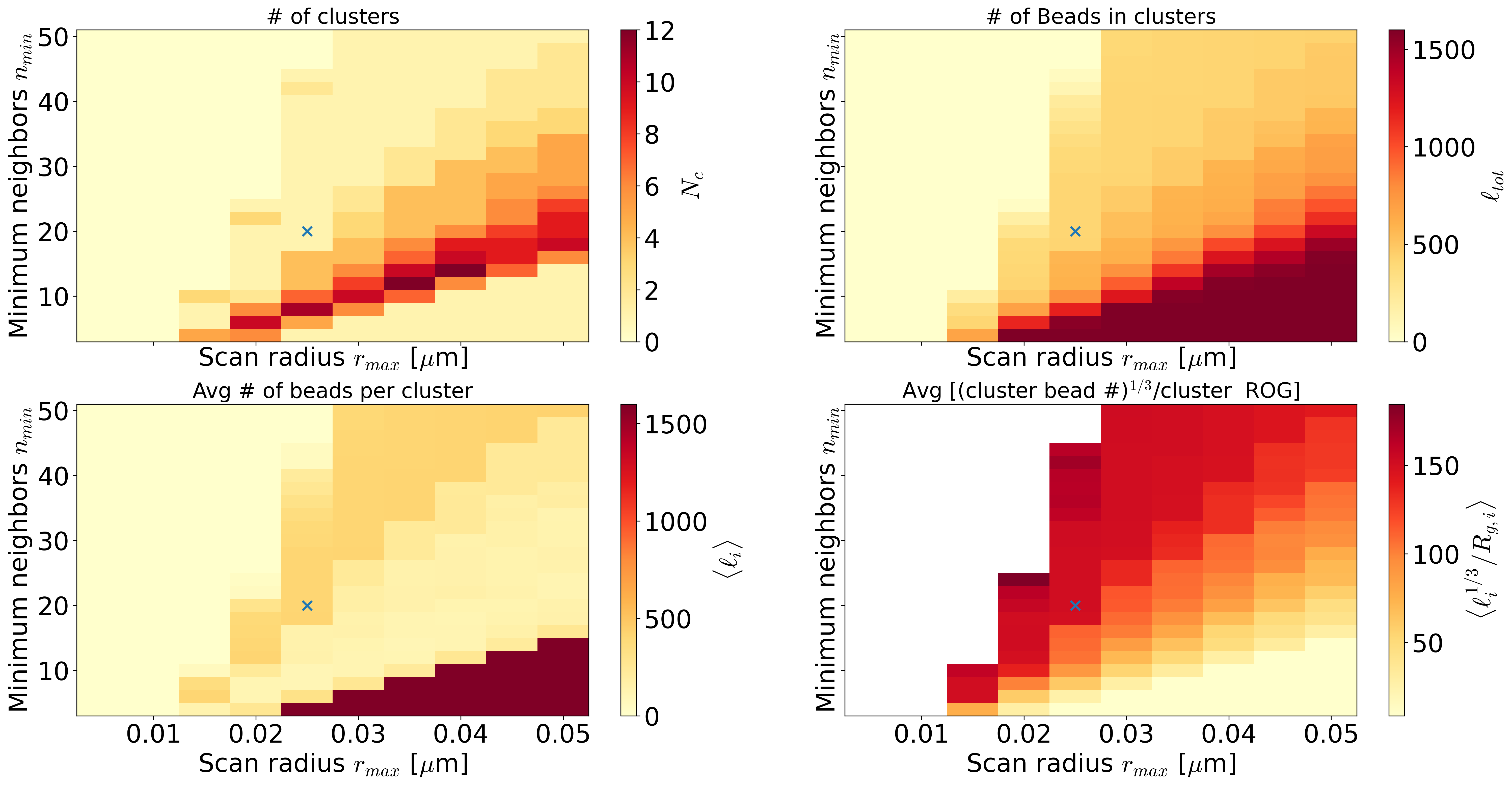}
    \caption{Steady-state average of identified cluster properties as a function
    of the DBSCAN parameters  $n_{min}$ (minimum neighbors) and  $r_{max}$ (scan
    radius). Results are shown for a system of $L_{sep} = 5\mu m$ and $K_e = 30
    \mu M^{-1}$. The number of clusters $N_c$ (top left), total number of beads
    in clusters $\ell_{tot}/b$ (top right), average number of beads per cluster
    $\ell_{tot}/bN_c$ (bottom left), and ratio between the cube root of cluster
    size and the cluster radius of gyration $\ell_i^{1/3}/ R_{g,i}$ (bottom
    right) are shown. Blue x shows the DBSCAN parameters used for analysis in
    this paper.} \label{sfig:dbscan_scan}
\end{figure}

DBSCAN efficiently clusters d-dimensional points based on density without  the
number of clusters needing to be specified \textit{a priori}. However, to
determine which points get classified into clusters, DBSCAN does require two
analysis parameters to be specified: the maximum distance for being considered a
neighboring point, $r_{max}$, and the minimum number of points to be considered
part of a cluster, $n_{min}$. To choose these parameters, we used snapshots from
simulations that clearly showed clustering and scanned over  $r_{max}$ and
$n_{min}$, plotting the average number of clusters, the total number of beads in
clusters, the average cluster size, and a shape characteristic
$\AVG{\ell_i^{1/3}/R_{g,i}}$, where $R_{g,i}$ is the radius of gyration of
cluster $i$, as shown in Fig.~\ref{sfig:dbscan_scan}.
According to Fig.~\ref{sfig:dbscan_scan}, decreasing $n_{min}$ increases the
total number of beads classified into clusters. However, this is because beads
that should be considered strands are considered to be parts of clusters as
$\AVG{\ell_i^{1/3}/R_{g,i}}$ decreases with $n_{min}$. Therefore, we would like
to maximize the average number of beads in clusters while minimizing the shape
characteristic. 

When inspecting the classified clusters, we noticed that with smaller $r_{max}$
values,  areas of high bead density would be classified as multiple clusters but
for larger $r_{max}$ and optimal $n_{min}$ values, many small transient clusters
would appear along the chain backbone from random fluctuations. This was
mitigated by favoring larger $r_{max}$ values but imposing a minimum on the
number of beads required to be considered a cluster. A final consideration was
ensuring that clusters would be robustly characterized from timestep to
timestep. Larger $n_{min}$ values maximized the shape characteristic but
medium-sized clusters tended be missed when their shapes fluctuated. This led us
to choose a smaller $n_{min}$ value for improved temporal tracking of clusters.

With these considerations, we chose values of $n_{min}=20$ and $r_{max} = .025
\mu\text{m}$. We also require that cluster have a minimum size of $40$.
Qualitatively, these parameters identify chain regions  that are in a compact
and roughly ellipsoidal configuration, intuitively corresponding to a droplet.

\subsection{Temporal tracking and history reconstruction}
\label{sub:appendix:cluster_def_tracking:temporal_track_history}
Tracking clusters in time poses significant challenges due to the dynamic nature
of the system. Unlike a static clustering problem, the number of clusters in our
system does not remain constant over time. Cluster composition also changes as
clusters merge or dissolve. Accurately capturing these events requires an
algorithm that can handle evolving, merging, and dissolving clusters.

This problem shares conceptual similarities with galaxy formation, where stars
and galactic structures merge and evolve over time. To study these processes in
astrophysics, the \verb|SUBLINK| algorithm was developed, providing a method of
capturing and analyzing the heredity of simulated galaxy structures. In our
study, we adapted the \verb|SUBLINK| algorithm \cite{Rodriguez-Gomez2015} by
implementing a modified Python version to quantify the timing and frequency of
both dissolving and merging events within polymeric systems. In essence, the
algorithm uses a merit function to determine the descendants of clusters in
time, identifying merging events of progenitor clusters.

The \verb|SUBLINK| algorithm tracks structural evolution by establishing
“progenitor-descendant” relationships between clusters across successive
simulation snapshots. At every timestep, clusters are compared based on their
composition $ g_i(t) = (\{p\}) $, where $p$ is a unique particle ID. A
descendant cluster is identified using a similarity threshold $|g_i(t) \cup
g_j(t+\Delta t)| / |g_i(t)| > \rho$. This approach constructs a “merger tree”
that captures the evolution and interactions of clusters, encompassing events
such as merging, splitting, and dissolving. By tracing these pathways, the
algorithm provides insight into the temporal dynamics of cluster formation and
disappearance within the system.

Galaxy-defining algorithms such as SubFind \cite{Springel2001} and FOF
\cite{Press1982} are designed to classify galactic structures with relatively
low uncertainty due to the sparse and distinct separations between particles.
Using these methods for connected polymeric systems would classify all beads as
a single cluster. Therefore, to adapt these approaches for our system,
modifications are necessary. While DBSCAN can be adjusted to mitigate this
issue, it introduces ambiguity for clusters near the threshold of cluster
definition.

To address these challenges, we modified the algorithm to allow a single cluster
to persist across multiple timesteps, even if it temporarily fails to meet the
DBSCAN criteria for times less than a chosen time window $w$. This adjustment
prevents clusters from being misclassified as distinct entities when they
fluctuate around the cluster criteria.  As a result, our approach enables more
robust and continuous tracking of clusters, even within the dynamic and
transient landscape of polymeric systems. In our analysis, we use the window
$w=10$ or $5$ seconds and a shared particle threshold of $\rho=0.4$.
Additionally, when two clusters merge, we consider the larger cluster to persist
while the smaller one is deemed to have ceased to exist at the time of merging.

\begin{algorithm}
\caption{Modified Find Descendants algorithm}
\begin{algorithmic}[1]
\STATE \textbf{Input:} Snapshots $\{S_t\}$ of clusters $\{C_i\}$, each with particle IDs $\{p\}$
\STATE \textbf{Parameters:} Threshold for minimum shared particle ratio $\rho$, Window of cluster persistence \(w\)
\STATE \textbf{Output:} Lineage tree capturing progenitor-descendant relationships, mergers, and disappearances 


\FOR{each snapshot $S_t$}
    \FOR{each cluster $C_i$ in $S_t$}
        \FOR{$i=1$ to $w$} 
            \FOR{each cluster $C_j$ in $S_{t+i}$}
                \STATE Calculate the number of shared particles between $C_i$ and $C_j$
                \STATE Compute overlap ratio:
                \[
                \text{overlap ratio} = \frac{\text{number of shared particles, } |\{p\}_i \cup \{p\}_j|}{\text{total particles in  \(C_i\), } |\{p\}_i|}
                \]
                \IF{overlap ratio $\geq \rho$}
                    \STATE Link $C_i$ to $C_j$ as a progenitor-descendant pair
                \ENDIF
            \ENDFOR
            \IF{At least one progenitor-descendant pair exists}
                \STATE \textbf{break}
            \ENDIF
        \ENDFOR
        \STATE Identify primary descendant and/or progenitor with the highest overlap ratio for $C_i$
    \ENDFOR
\ENDFOR

\FOR{each cluster}
    \STATE Build lineage tree by recursively linking progenitors and descendants (\cite{Rodriguez-Gomez2015})
\ENDFOR

\STATE \textbf{Special Event Handling:}
\FOR{each cluster}
    \IF{multiple clusters in $S_t$ have the same primary descendant in $S_{t+1}$}
        \STATE Mark event as a \textbf{merger}
    \ENDIF
    \IF{a cluster in $S_t$ has no descendant in $S_{t+1}$}
        \STATE Mark event as \textbf{dissolved}
    \ENDIF
\ENDFOR

\STATE Output lineage tree with progenitor-descendant links, mergers, and disappearances.

\end{algorithmic}
\end{algorithm}

\begin{figure}
    \centering
    \includegraphics[width=1.1\linewidth]{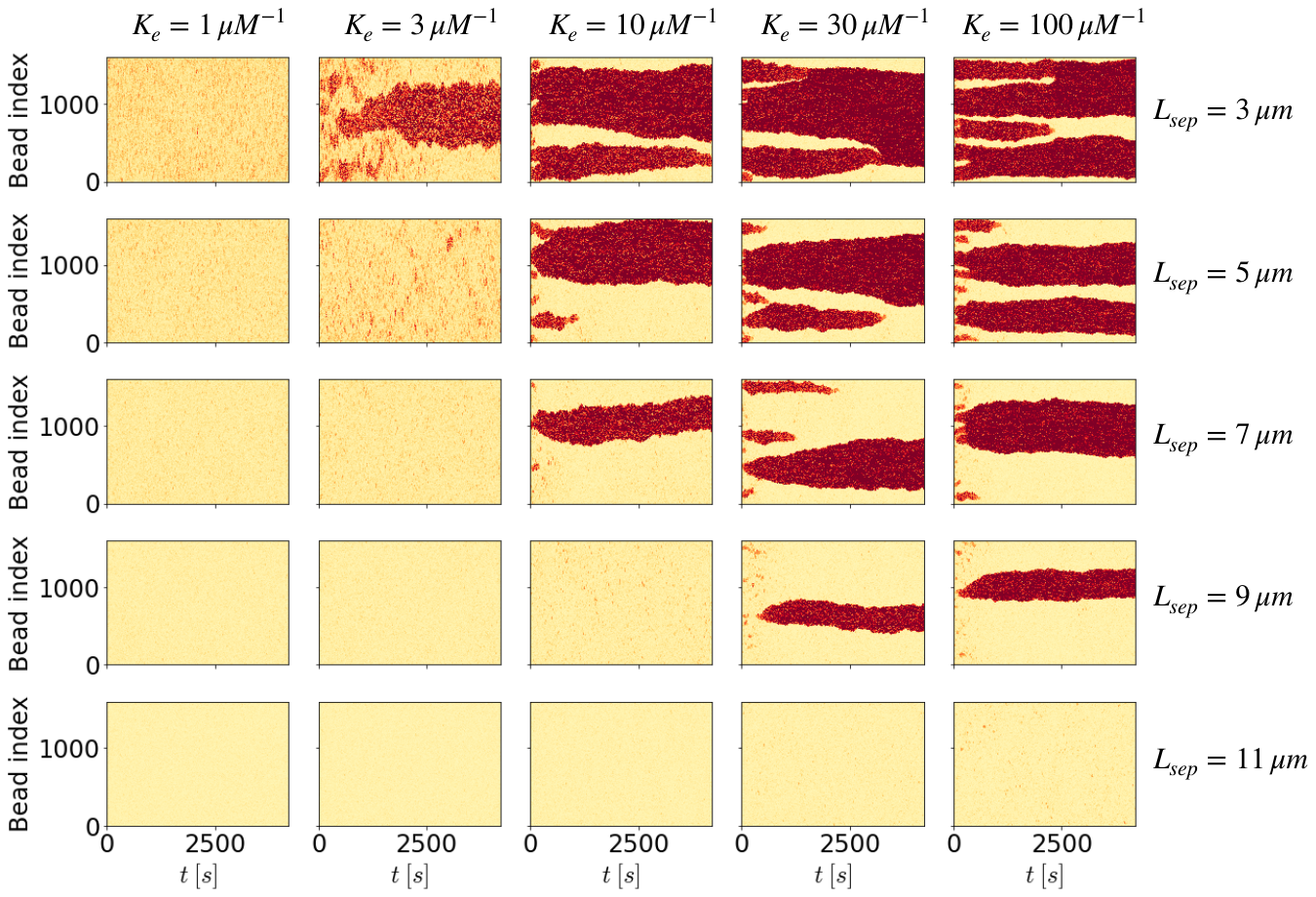}
    \caption{Summed contact probability per bead over time showing which beads
    exist in regions of high density (red).  Each panel is one realization of a
    simulation from Fig.~\ref{figs:fig2}(B). Columns show simulations with the
    same tail binding affinity $K_e$ while rows show simulations with the same
    chain end-to-end separation $L_{sep}$.}
    \label{figs:fig2_supp_contact_kymo_mod}
\end{figure}

\begin{figure}
    \centering
    \includegraphics[width=0.9\linewidth]{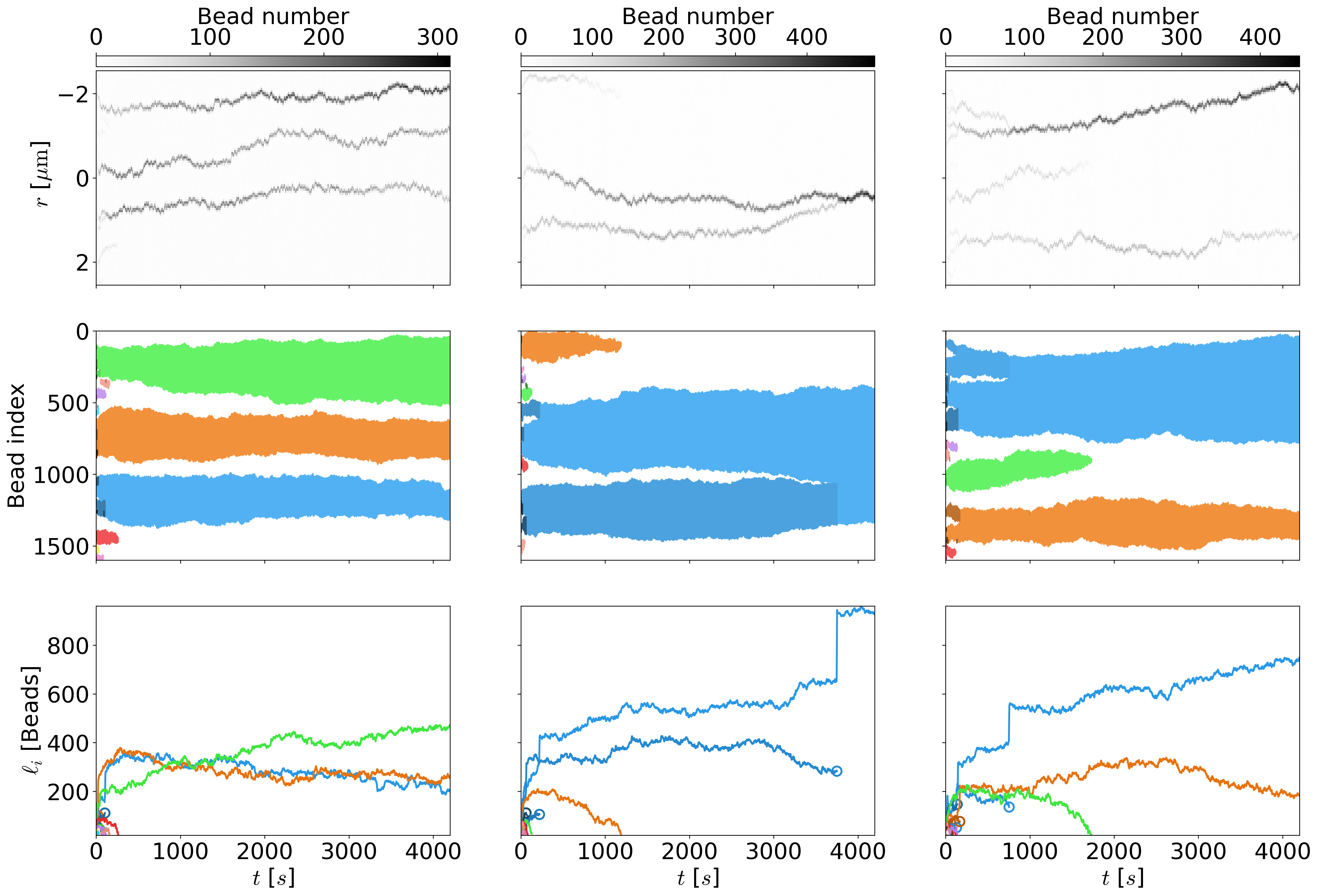}
      \caption{Visualization of cluster dynamics from individual realizations of
      a simulation with parameters $L_{sep} = 5\mu m$ and $K_e = 30 \mu M^{-1}$.
      (Top row) Histogram of bead number along the chain end separation axis.
      Ends of the chain are located at $\pm 2.5$. Clusters shown in the bead
      index space over time. (Middle row) Filled regions of the same color
      indicate bead indices belonging to the same cluster, with darker shades
      representing earlier time points in the cluster's genealogy. (Bottom row)
      Cluster sizes from the middle row as a function of time, with merging
      events marked by circles.
  }
    \label{figs:fig5_supp_tracking_multiple}
\end{figure}

\end{document}